\documentclass[aps,pra,twocolumn,showpacs,floatfix]{revtex4}
\usepackage{graphicx,amsmath,amssymb}
\usepackage{times}
\usepackage{txfonts}

\begin{document}
  
  \title{Excited Thomas-Efimov levels in ultracold gases}
  
  \author{Mark D.~Lee}
  \affiliation{Clarendon Laboratory, Department of Physics,
    University of Oxford, Parks Road, Oxford, OX1 3PU, UK}
  
  \author{Thorsten K{\"o}hler}
  \affiliation{Department of Physics and Astronomy, University College London, 
    Gower Street, London, WC1E 6BT, UK}
  
  \author{Paul S.~Julienne}
  \affiliation{Atomic Physics Division, National Institute of Standards and 
    Technology, 100 Bureau Drive Stop 8423, Gaithers\-burg, Maryland 
    20899-8423}
  
  \begin{abstract}
    Since the early days of quantum physics, the complex behavior of three 
    interacting particles has been the subject of numerous experimental and 
    theoretical studies. In a recent Letter to Nature, Kraemer \emph{et al.} 
    [Nature (London) \textbf{440}, 315 (2006)] report on experimental 
    ``evidence for Efimov quantum states'' in an ultracold gas of cesium 
    atoms. Such quantum states refer to an infinite series of energy levels of 
    three identical Bose particles, accumulating at the threshold for 
    dissociation as the scattering length of each pair is tuned to infinity. 
    Whereas the existence of a single Efimov state has been predicted for 
    three helium atoms, earlier experimental studies concluded that this 
    elusive state had not been found. In this paper we show by an intuitive
    argument and full numerical calculations that the helium and cesium 
    experiments actually provide evidence of the same, ground state of this 
    trimer spectrum, which the helium experimentalists and pioneering 
    theoretical studies had not associated with Efimov's effect. Unlike the 
    helium trimer, the observed $^{133}$Cs$_3$ resonance refers to a 
    Borromean molecular state. We discuss how as yet unobserved, excited 
    Efimov quantum states might be detected in ultracold gases of $^{85}$Rb 
    and of $^{133}$Cs at magnetic field strengths in the vicinity of 0.08\,T
    (800\,G). 
  \end{abstract}
  
  \date{\today}
  \maketitle
  
  \section{Introduction}
  
\begin{figure}[htb]
    \includegraphics[width=\columnwidth,clip]{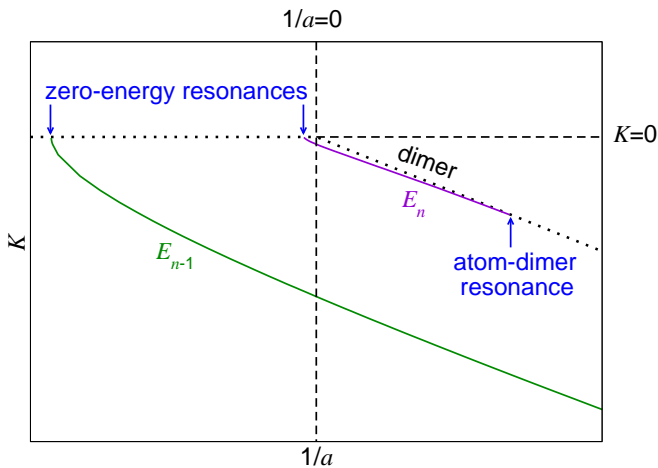}
    \caption{(Color online) Efimov plot \cite{Efimov1970a} illustrating the 
      wave number $K=-(m|E_n|/\hbar^2)^{1/2}$ ($n=1,2,3,\ldots$) indicated by 
      solid curves vs.~the inverse scattering length, $1/a$. Here $m$ is the 
      single particle mass and $E_n$ is the energy of the $n$th three-particle
      level. The wave number labeled dimer (dotted line) is associated with 
      the two-body bound state emerging in the limit $1/a\to 0$ 
      (i.e.~$|a|\to\infty$) at the vertical dashed line. In accordance with 
      Ref.~\cite{Efimov1970a}, its energy is universal, 
      i.e.~$E_\mathrm{b}=-\hbar^2/(ma^2)$. The limits $1/a\to\pm\infty$ of the 
      horizontal axis refer to zeros of the scattering length. Ideally, each 
      Efimov level emerges in a three-body zero-energy resonance at negative 
      $a$ and eventually vanishes into the continuum leading to an atom-dimer 
      resonance at positive $a$. The ground state of the Thomas-Efimov 
      spectrum ($n=1$) is not displayed.}
    \label{fig:Efimovplot}
  \end{figure}

  Efimov's effect \cite{Efimov1970a,Efimov1971a,Jensen2004a} refers to a 
  scenario in which three identical Bose particles interact via weak pair 
  potentials, each supporting only a single bound state when the $s$-wave 
  scattering length $a$ is positive. None of the pairs is bound when $a$ is 
  negative, while at the intersection of the two regimes the scattering length 
  has a singularity. As early as 1935 Thomas \cite{Thomas1935a} showed that 
  under these conditions three particles can be tightly bound, despite their 
  weak binary interactions. Thirty five years later, Efimov \cite{Efimov1970a} 
  discovered a striking extension of Thomas' effect, predicting the existence 
  of an infinite series of excited three-body energy levels in the limit 
  $|a|\to\infty$. Each Efimov state emerges when $a$ is negative at the 
  threshold for dissociation into three free particles and vanishes eventually 
  via decay into a two-body bound state and a third free particle as $a$ is 
  increased to positive values across the singularity. Such an ideal universal 
  scaling behavior of three-particle energy levels is illustrated in 
  Fig.~\ref{fig:Efimovplot}. Two of the levels (labeled by their degree of 
  excitation, $n=1,2,3,\ldots$) are visible but ideally they all have the same 
  functional form as $a$ is varied, the full extent of which is displayed only 
  for the $E_n$ level.
  
  Strict universality in a three-body system presupposes a large scattering 
  length whose modulus by far exceeds all the other length scales set by the 
  pair potentials. This specific requirement on the two-particle interactions 
  in combination with the wide spread of energy scales in 
  Fig.~\ref{fig:Efimovplot} have for a long time prevented any generally 
  accepted discovery of Efimov states in nature. Experimental realizations of 
  Efimov's scenario have been suggested, for instance, for trimer molecules 
  consisting of three $^4$He atoms \cite{Lim1977a}, as well as for alkali 
  atoms in ultracold gases with magnetically tunable interactions (see 
  Refs.~\cite{Nielsen1999a,Esry1999a,Kraemer2006a} and references therein). 
  Helium trimers well match the assumptions underlying Thomas' and Efimov's 
  effects, as for each atom pair there exists only a single weakly bound dimer 
  state giving rise to the large positive scattering length of about 
  \cite{Grisenti2000a} $200\,a_0$ ($a_0$ being the Bohr radius). The 
  inter-atomic interactions are inaccessible to external manipulation but are 
  sufficiently weak to give rise to two trimer bound states of the 
  Thomas-Efimov spectrum. 
  
  The excited $^4$He$_3$ $E_2$ level follows the qualitative trends indicated 
  in Fig.~\ref{fig:Efimovplot} when the pair potential is scaled to mimic a 
  hypothetical tuning of $a$. In particular, it vanishes into the atom-dimer 
  continuum as the pairwise attraction is increased 
  \cite{Cornelius1986a,Esry1996a} and has therefore been referred to as a 
  genuine Efimov state. By contrast, a rigorous variational estimate 
  \cite{Bruch1973a} reveals that the binding energy of the ground state of the 
  Thomas-Efimov spectrum (not shown in Fig.~\ref{fig:Efimovplot}) violates 
  the global universal scaling behavior of Fig.~\ref{fig:Efimovplot}, as it 
  never intersects with the two-body level. Similar deviations from three-body 
  universality, as $1/a$ is hypothetically increased, have been discussed in 
  Ref.~\cite{Efimov1981a}, in the context of the three-nucleon problem. In 
  particular, several publications 
  \cite{Cornelius1986a,Esry1996a,Schollkopf1994a,Bruhl2005a} had not 
  associated the experimentally observed \cite{Schollkopf1994a,Bruhl2005a} 
  $^4$He$_3$ ground state with Efimov's effect. In the limit of zero 
  interaction, i.e.~$1/a\to-\infty$ in Fig.~\ref{fig:Efimovplot}, however, 
  this ground trimer level must vanish in a three-body zero-energy resonance. 
  
  In this paper we show that recent experiments in ultracold gases of 
  $^{133}$Cs atoms \cite{Kraemer2006a} also provide evidence for the ground 
  trimer level of the Thomas-Efimov spectrum via observation of its associated 
  zero-energy resonance. We investigate several Efimov spectra of the 
  experimentally relevant alkali atomic species $^{133}$Cs and $^{85}$Rb in 
  the vicinity of singularities of the diatomic scattering length, realized by 
  magnetically tunable Fesh\-bach resonances 
  \cite{Stwalley1976a,Tiesinga1993a,Inouye1998a}. This involves predictions of 
  the energies of metastable trimer levels as well as numerical calculations 
  of three-body recombination loss-rate constants 
  \cite{Nielsen1999a,Esry1999a}, motivated by the observations reported in 
  Ref.~\cite{Kraemer2006a}. Our studies indicate that ultracold gases in the 
  presence of broad, entrance-channel dominated diatomic zero-energy 
  resonances and negative background-scattering lengths \cite{Koehler2006a} 
  may be best suited for the detection of as yet unobserved excited trimer 
  levels of the Thomas-Efimov spectrum \cite{Stoll2005a}.
  
  The paper is organized as follows: In Section~\ref{sec:3bodyCs} we briefly 
  discuss the technique of detecting three-body zero-energy resonances 
  \cite{Nielsen1999a,Esry1999a,Kraemer2006a} in ultracold gases of $^{133}$Cs 
  at low magnetic-field strengths. Based on quantitative calculations of 
  associated three-body recombination loss-rate constants, we interpret the 
  measurements of Ref.~\cite{Kraemer2006a} in terms of their relation to 
  Efimov's effect \cite{Efimov1970a,Efimov1971a}. In this context, we discuss 
  crucial shortcomings of purely universal treatments of near-resonant 
  three-body recombination into comparatively tightly bound target-dimer 
  states at negative as well as small positive scattering lengths 
  \cite{Braaten2001a,Braaten2006a,Kraemer2006a}. Using an intuitive argument, 
  we illustrate our finding that the experimentally observed three-body 
  zero-energy resonance \cite{Kraemer2006a} refers to the ground state of the 
  Thomas-Efimov spectrum. In Section~\ref{sec:excitedEfimov} we predict 
  magnetic-field strengths associated with excited state three-body 
  zero-energy resonances in ultracold $^{133}$Cs gases in the vicinity of 
  800\,G. We discuss prospects for their detection in comparison with a 
  previous suggestion for experiments using $^{85}$Rb \cite{Stoll2005a}. 
  Section~\ref{sec:conclusions} summarizes our conclusions. Finally, the 
  appendix provides a detailed description of our general method 
  \cite{Smirne2006a} for calculating resonance-enhanced three-body 
  recombination loss-rate constants including the regimes of negative as well 
  as small positive scattering lengths. 
  
  \section{Resonance-enhanced three-body recombination in cesium gases}
  \label{sec:3bodyCs}
  
  Whereas the helium studies of Refs.~\cite{Schollkopf1994a,Bruhl2005a} relied 
  upon direct observation of $^4$He$_3$ in a molecular beam, the existence of 
  trimer levels in ultracold gases \cite{Kraemer2006a} has been inferred from 
  three-body zero-energy resonances as illustrated in 
  Fig.~\ref{fig:Efimovplot}. Their signatures manifest themselves, for 
  example, in a resonant enhancement of three-body recombination loss 
  \cite{Nielsen1999a,Esry1999a}. In such processes three initially free atoms 
  collide to form a dimer, while the binding energy is transferred to the 
  relative motion of the third atom with respect to the bound pair. Their 
  final kinetic energies are usually high enough for all three colliding 
  particles to leave an atom trap confining an ultracold gas, whose atom 
  number $N$ thus decays in accordance with the rate equation
  \begin{equation}
    \label{rateequation}
    \dot{N}=-K_3\langle n^2\rangle N. 
  \end{equation}
  Here $\langle n^2\rangle$ is the mean square density and $K_3$ is the 
  three-body recombination loss-rate constant.
  
  \subsection{Ultracold collisions of cesium atoms at low magnetic-field 
  strengths}
  \label{subsec:lowfield}
  
  \begin{figure}[htb]
    \includegraphics[width=\columnwidth,clip]{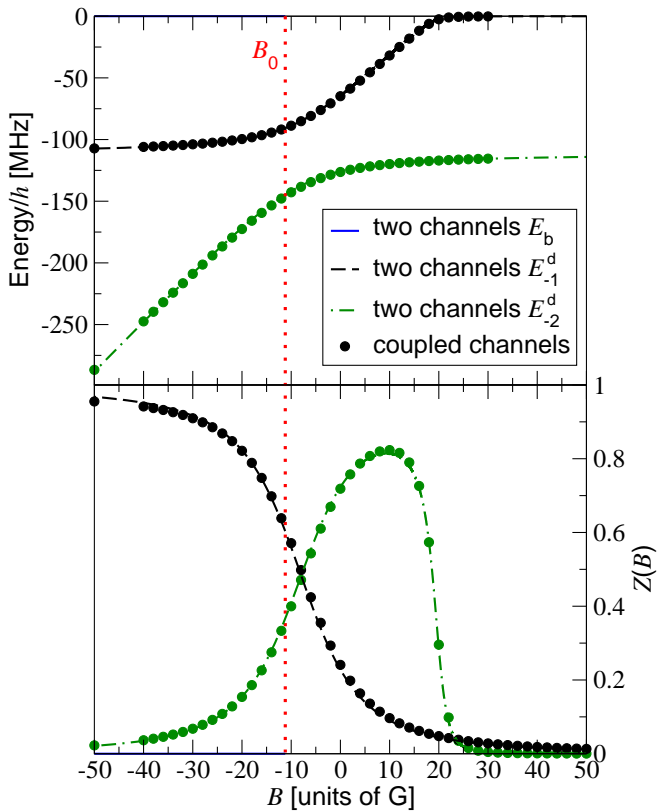}
    \caption{(Color online)
    Comparison between full coupled-channels (circles) and two-channel 
    (curves) calculations of the $^{133}$Cs 6$s$-dimer energies (upper panel)
    \cite{Tiesinga1993a,Moerdijk1995a,Chin2004a}, as well as their associated 
    closed-channel admixtures \cite{Koehler2006a}, $Z(B)$ (lower panel), 
    vs.~the magnetic-field strength, $B$. For each magnetic-field strength 
    zero energy is chosen to coincide with the scattering threshold. The 
    resonance position, $B_0$, is indicated by the dotted line. Curves labeled 
    $E^\mathrm{d}_{-1}$ and $E^\mathrm{d}_{-2}$ refer to the two highest 
    excited vibrational bound states for $B>B_0$ and the levels they 
    adiabatically correlate with at $B<B_0$, whereas the Fesh\-bach molecular 
    energy $E_\mathrm{b}$ existing for $B<B_0$ is not resolved in the upper 
    panel. Its closed-channel admixture in the lower panel is negligible.}
    \label{fig:133CsEbvsB}
  \end{figure}
  
  Observation of atom loss was used in Ref.~\cite{Kraemer2006a} to track down 
  a three-body zero-energy resonance in gases of $^{133}$Cs prepared in the 
  $(F=3,m_F=+3)$ Zeeman ground state at temperatures $\geq10\,$nK. Here $F$ 
  refers to the total atomic angular momentum quantum number, and $m_F$ 
  indicates the associated projection with respect to an applied, spatially 
  homogeneous magnetic field of strength $B$. This experimental setup 
  \cite{Kraemer2006a} allows magnetic tuning of the $s$-wave scattering length 
  $a$ to mimic the variation of inter-particle interactions in Efimov's 
  scenario \cite{Efimov1970a,Efimov1971a}, taking advantage of the occurrence 
  of a diatomic zero-energy resonance in cesium collisions at low fields.

  In the vicinity of any diatomic zero-energy resonance \cite{Taylor1972a}, 
  the scattering length assumes all values between $-\infty$ and $+\infty$. 
  The associated strong magnetic-field variation $a(B)$ can usually be 
  described, to an excellent level of accuracy, by the following general 
  formula \cite{Moerdijk1995a}:
  \begin{equation}
     a(B)=a_\mathrm{bg}\left(1-\frac{\Delta B}{B-B_0}\right).
     \label{aofB}
  \end{equation}   
  Here $B_0$ is the position where $a$ has a singularity, $a_{\rm bg}$ is the 
  background scattering length observed asymptotically far from $B_0$, while 
  the resonance width, $\Delta B$, refers to the distance between $B_0$ and 
  the zero-crossing point of the scattering length, i.e.~$a(B_0+\Delta B)=0$. 
  
  In the particular case of $^{133}$Cs atoms prepared in the $(F=3,m_F=+3)$ 
  state near zero field strength, the strong $B$-dependence of the scattering 
  length is driven by a diatomic zero-energy resonance in the 
  mirror-spin-channel of atoms in the excited $(F=3,m_F=-3)$ Zeeman state. 
  This scenario can be formally described \cite{Vogels1998a} by a negative 
  resonance position that we calculate at $B_0=-11.2\,$G (1\,G=$10^{-4}\,$T) 
  using the complete two-body Hamiltonian model of Ref.~\cite{Chin2004a}, 
  which accurately represents the positions of all known diatomic resonances 
  in cesium collisions. An associated vibrational molecular energy spectrum is 
  illustrated in the upper panel of Fig.~\ref{fig:133CsEbvsB} for isotropic 
  $6s$-dimer states \cite{Tiesinga1993a,Moerdijk1995a,Chin2004a}, whose 
  highest excited, Fesh\-bach molecular, level $E_\mathrm{b}$ 
  \cite{Koehler2006a} (unresolved in Fig.~\ref{fig:133CsEbvsB}) causes the
  zero-energy resonance at $B_0$. The large, positive background-scattering 
  length of $a_\mathrm{bg}=1720\,a_0$, in combination with the resonance width 
  of $\Delta B=27.5\,$G, allow $a(B)$ to be varied between about $-2500\,a_0$ 
  and zero within a field range between $B=0$ and about $17\,$G. Beyond its 
  zero-crossing point the scattering length rises to $a_\mathrm{bg}$, as shown 
  in the inset of the upper panel of Fig.~\ref{fig:133CsK3vsB}.

  \subsection{Three-body zero-energy resonances}

  \begin{figure}[htb]
    \includegraphics[width=\columnwidth,clip]{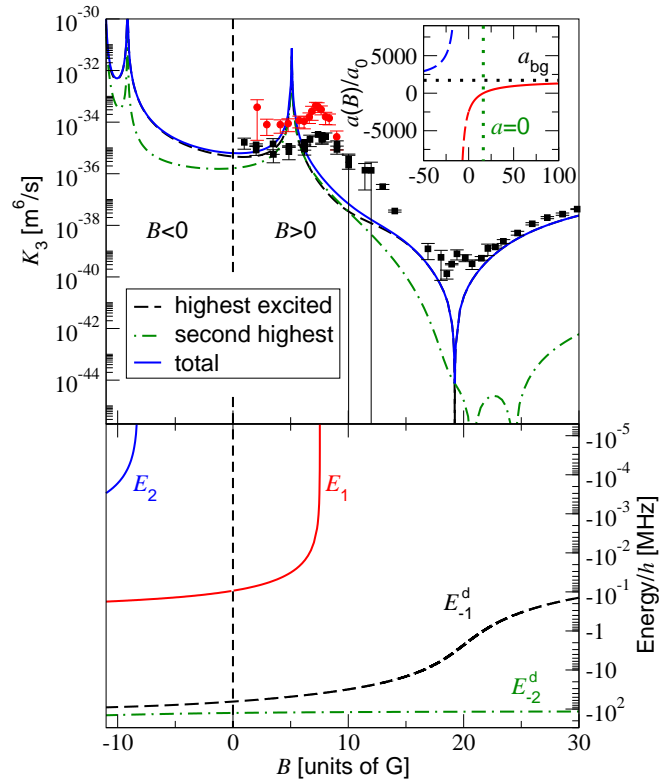}
    \caption{(Color online) 
    Calculated three-body recombination loss-rate constants $K_3$ (upper 
    panel) and energies of metastable trimer levels of $^{133}$Cs atoms
    labeled $E_1$ and $E_2$ (lower panel) vs.~the magnetic-field strength $B$.
    For each magnetic-field strength zero energy coincides with the scattering 
    threshold for three asymptotically free atoms. Only ranges of $B>0$ are 
    accessible to experiments with atoms in the Zeeman ground state, whereas 
    $B<0$ would be accessible to experiments with the $(F=3,m_F=-3)$ level. 
    Measurements of $K_3$ \cite{Kraemer2006a,Weber2003a} performed at 10\,nK 
    and about 200\,nK are indicated by circles and squares, respectively. The 
    dashed and dot-dashed curves labeled $E^\mathrm{d}_{-1}$ and 
    $E^\mathrm{d}_{-2}$ in the lower panel refer to the $6s$-dimer levels 
    \cite{Tiesinga1993a,Moerdijk1995a,Chin2004a} of Fig.~\ref{fig:133CsEbvsB}, 
    constituting the target states included in our calculations. Dashed and 
    dot-dashed curves in the upper panel indicate their associated 
    contributions to the total three-body recombination loss-rate constant 
    (solid curve). The inset of the upper panel illustrates the variation of 
    the scattering length with magnetic field.}
    \label{fig:133CsK3vsB}
  \end{figure}

  Based on numerically exact three-body calculations without fit parameters 
  using the method of Ref.~\cite{Smirne2006a}, Fig.~\ref{fig:133CsK3vsB} shows 
  the $B$-dependence of $K_3$ as well as the metastable $^{133}$Cs$_3$ level 
  causing a three-body zero-energy resonance observed in 
  Ref.~\cite{Kraemer2006a} at 7.2\,G. Our estimates of its position rely upon 
  two independent methods: The first approach \cite{Stoll2005a} follows 
  Thomas' and Efimov's original treatments and accounts only for the weakly 
  bound Fesh\-bach molecular dimer level \cite{Koehler2006a}, which causes the 
  diatomic zero-energy resonance at $B_0$ in the mirror-spin-channel. This 
  leads to a zero crossing of the energy of a trimer state at 7.5\,G (solid 
  curve labeled $E_1$ in Fig.~\ref{fig:133CsK3vsB}). 
  
  Our second approach consists in directly determining the peak positions in 
  the calculated magnetic-field dependent three-body recombination loss-rate 
  constants \cite{Smirne2006a}. In addition to the Fesh\-bach molecule, these 
  calculations account for the next two highest excited $6s$-dimer levels 
  \cite{Chin2004a} below the dimer dissociation limit, including their 
  admixtures from different atomic Zeeman states \cite{Koehler2006a}, as 
  discussed in the appendix. The associated pair potentials are based on a 
  parametric two-channel representation of the diatomic low-energy spectra 
  resulting solely from coupled-channels calculations. The one approximation 
  we make in the two-body coupled-channels bound-state calculations is to 
  neglect the relatively weak coupling of the 6$s$-levels to nearby dimer 
  levels of predominantly $d$-wave symmetry. We plan future calculations in 
  which we include such levels. Our exact solutions of the three-body 
  Schr\"odinger equation have been performed at zero kinetic energy of the 
  three initially free colliding atoms and lead to a resonance peak in $K_3$ 
  at 5.1\,G. We note that the second approach could also be used to calculate
  the Thomas-Efimov trimer levels of Fig.~\ref{fig:133CsK3vsB} in terms of 
  resonant enhancements of atom-dimer collision-rate constants 
  \cite{Smirne2006a}, rather than the stable bound states of the first 
  approach. In such a more precise treatment, the atom-dimer resonances become 
  degenerate with the zero-energy scattering threshold exactly at those
  magnetic-field strengths where three-body zero-energy resonances occur in 
  the recombination loss-rate constant $K_3$.
  
  Relative to the width of the diatomic zero-energy resonance the deviations 
  between the calculated and observed peak positions are less than 8\,\% for 
  both methods without any adjustments to the measured $K_3$. We note that 
  reducing the temperature of the cesium gas tends to shift the peak towards 
  lower fields \cite{Naegerl2006a}. The comparison between both methods and 
  the experiments indicates that the three-body zero-energy resonance 
  positions in the upper panel of Fig.~\ref{fig:133CsK3vsB} are largely 
  insensitive to the details of the diatomic molecular target levels. This 
  behavior is physically sensible, as these positions depend only on the 
  metastable Thomas-Efimov levels close to the three-body dissociation 
  threshold in the lower panel.
  
  \subsection{Incompleteness of universal approaches to three-body
    recombination at negative scattering lengths}
  
  As opposed to universal fitting procedures \cite{Braaten2001a,Braaten2006a}, 
  our calculations also correctly incorporate the two- and three-body physics 
  in the vicinity of the minimum of $K_3$ (just below 20\,G in the upper panel 
  of Fig.~\ref{fig:133CsK3vsB}) and fully recover its measured position. While 
  the overall calculated magnitudes are in good agreement with experiment 
  throughout, we observe a strong sensitivity of $K_3$ to the $B$-dependent 
  closed-channel admixtures \cite{Koehler2006a} of the diatomic target states 
  illustrated in the lower panel of Fig.~\ref{fig:133CsEbvsB}. This 
  sensitivity at negative as well as small positive scattering lengths is to 
  be expected because the formation of a dimer from initially free atoms 
  should be suppressed when the atoms have to change their Zeeman states 
  during this process. These findings imply that $K_3$ depends on $B$ not only 
  through $a(B)$ but also through the strongly varying properties of the 
  comparatively tightly bound diatomic molecular target levels of 
  Fig.~\ref{fig:133CsEbvsB}. Consequently, any fit analysis of the measured 
  data in Fig.~\ref{fig:133CsK3vsB} incorporating the $B$-dependence of $K_3$ 
  only through the scattering length is incomplete, and could therefore lead 
  to unreliable conclusions about three-body universality in the system.

  \subsection{Interpretation of the cesium experiments}

  \begin{figure}[htb]
    \includegraphics[width=\columnwidth,clip]{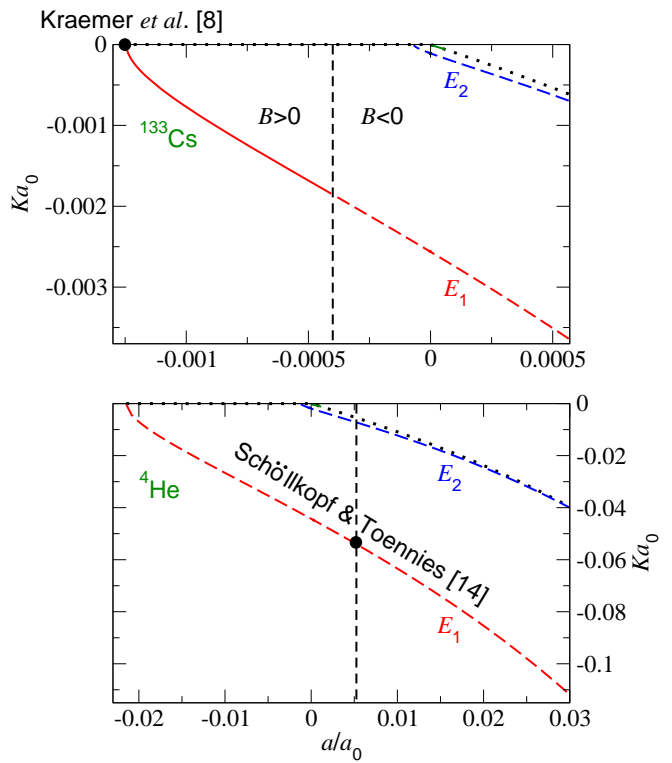}
    \caption{(Color online) 
    Comparison between Efimov plots associated with $^{133}$Cs in the vicinity 
    of zero field (upper panel) and $^4$He (lower panel). The filled circles 
    refer to the degree of excitation ($n=1$) of the trimer states detected in 
    the experiments of Refs.~\cite{Schollkopf1994a,Kraemer2006a} at the 
    inverse scattering lengths indicated on the horizontal axis. Dotted curves 
    indicate energies of dimer levels causing diatomic zero-energy resonances 
    at $1/a=0$. The range of scattering lengths in the upper panel extends 
    from the observed \cite{Kraemer2006a} three-body zero-energy resonance 
    position to $a=a_\mathrm{bg}$ at field strengths $B<0$. In accordance with 
    Refs.~\cite{Bruch1973a,Efimov1981a}, the curves labeled $E_1$ illustrate 
    the increasing separation of the trimer ground level from the dimer energy 
    as $1/a$ is increased. By contrast, the $E_2$ level in the lower panel 
    approaches the atom-dimer threshold. It eventually gets overrun by the 
    dimer energy causing an atom-dimer resonance as the potential is scaled to 
    enhance the pairwise attraction \cite{Cornelius1986a,Esry1996a}. For this 
    reason, only the excited $E_2$ level has been referred to in 
    Refs.~\cite{Lim1977a,Cornelius1986a,Esry1996a,Bruhl2005a} as a genuine 
    Efimov state.}
    \label{fig:compHeCs}
  \end{figure}
  
  Figure~\ref{fig:133CsK3vsB} shows that according to our calculations the 
  resonance peak experimentally observed at 7.2\,G \cite{Kraemer2006a} is 
  associated with the ground trimer level of the Thomas-Efimov spectrum, 
  $E_1$. As opposed to the peak positions and magnitudes of $K_3$, the labels 
  attached to the trimer levels in the lower panel are independent of the 
  detailed physics included in the approach. All that is needed for this 
  classification is the knowledge that the observed resonance 
  \cite{Kraemer2006a} is the first to appear on the low-field side of the zero 
  crossing of the scattering length. This is immediately evident from the 
  $^{133}$Cs Efimov plot in the upper panel of Fig.~\ref{fig:compHeCs}, in 
  which the field strength associated with $a=0$ corresponds to the limit 
  $1/a\to-\infty$. As $B$ is decreased the inverse scattering length $1/a(B)$ 
  increases. Zero magnetic-field strength is indicated by the vertical dashed 
  line. The trimer energy spectrum beyond this point is indicated by dashed 
  curves in the upper panel of Fig.~\ref{fig:compHeCs} and might be accessed 
  in gases of $^{133}$Cs atoms prepared in the excited $(F=3,m_F=-3)$ Zeeman 
  state. 
  
  For comparison, the lower panel of Fig.~\ref{fig:compHeCs} shows the Efimov 
  plot associated with $^4$He$_3$ \cite{Schollkopf1994a,Bruhl2005a}. Both the 
  cesium \cite{Kraemer2006a} and helium \cite{Schollkopf1994a,Bruhl2005a} 
  experiments provide evidence for the same level of the Thomas-Efimov 
  spectrum. Whereas the $^4$He$_3$ molecule is stable with respect to 
  spontaneous dissociation, the $^{133}$Cs$_3$ state can decay into an atom 
  and, e.g., one of the dimer levels in the lower panel of 
  Fig.~\ref{fig:133CsK3vsB} in accordance with energy conservation. We note 
  in this context that all stable $^{133}$Cs$_3$ molecular levels are below 
  the diatomic ground-state energy, whose typical modulus for alkali systems 
  exceeds the largest $h\times 100\,$MHz energy scale considered in this paper 
  by six orders of magnitude. Consequently, the ground state of the 
  Thomas-Efimov spectrum of cesium does not coincide with the stable 
  molecular ground state of $^{133}$Cs$_3$. The resonance signature of the 
  observed \cite{Kraemer2006a} metastable Thomas-Efimov cesium trimer, 
  however, has been detected at a negative scattering length. To our 
  knowledge, this measurement therefore provides the first experimental 
  evidence for the existence of a Borromean \cite{Zhukov1993a} molecular 
  state, i.e.~a three-body energy level, with a finite width in the present 
  context, in the absence of any diatomic levels energetically above it. Among 
  the exotic three-atom bound molecular states of helium, including the 
  isotope mixture $^4$He$_2$$^3$He, for instance, at least one atom pair, 
  namely $^4$He$_2$, is bound \cite{Schollkopf1994a}. Accordingly, only a 
  ``pseudo-Borromean'' four-atom complex, $^4$He$_2$$^3$He$_2$, has been 
  observed in molecular-beam experiments \cite{Kalinin2005a}. 
  
  \section{Excited Thomas-Efimov trimer levels}
  \label{sec:excitedEfimov}
 
  The prospect of an experimental verification of Efimov's effect depends on 
  the criterion for its discovery, since a measurement of an infinite number 
  of trimer levels with increasingly loose bonds is impossible. Numerous 
  pioneering studies, such as Refs.~\cite{Lim1977a,Cornelius1986a,Esry1996a}, 
  therefore agreed that the detection of a single Efimov state would be 
  sufficient. References~\cite{Braaten2003a,Braaten2006a} suggest that the 
  $^4$He$_3$ ground state should be referred to as the first example of an 
  Efimov state in nature, as the modulus of the diatomic scattering length by 
  far exceeds all the other length scales set by the pair potentials 
  \cite{Grisenti2000a}. Several studies summarized in 
  Refs.~\cite{Braaten2003a,Stoll2005a} have shown that the $^4$He$_3$ molecule 
  is indeed universal, i.e.~both levels are accurately determined by the 
  scattering length in addition to a single independent parameter 
  \cite{Efimov1970a,Braaten2006a}. Such a parameter could be provided, e.g., 
  by recent experimental studies \cite{Bruhl2005a} showing that the bond 
  lengths of the observed $^4$He$_3$ molecules \cite{Schollkopf1994a} are 
  consistent with predictions for the ground state. Several theoretical 
  studies suggest that the ground-state triton $^3$H nucleus
  \cite{Oliphant1934a} might also be described using universal treatments 
  (see, e.g., Refs.~\cite{Efimov1981a,Adhikari1982a,Efimov1991a,Efimov1993a,Braaten2003b}). By contrast, a recent publication \cite{Braaten2006c} concludes 
  that the discovery of the cesium three-body zero-energy resonance 
  \cite{Kraemer2006a} provided the first evidence for the existence of 
  Efimov's effect in nature. This assessment is based on a fit analysis of the 
  measurements universally relating the resonance position of 7.2\,G to the 
  field strength of the minimum of $K_3$ just below 20\,G in 
  Fig.~\ref{fig:133CsK3vsB}. Such a procedure relies upon the assumption that 
  universality is preserved across the zero of scattering length and that all 
  $B$-dependence of $K_3(B)$ is included in $a(B)$, whereas this might be 
  strictly justified at most in the limit $a\to+\infty$.

  \subsection{Three-body zero-energy resonances of $^{133}$Cs at high field 
    strengths in the vicinity of 800\,G}
  \label{subsec:cesium800G}
  
  Besides the existing evidence for the ground state of the Thomas-Efimov 
  spectrum in, for instance, $^4$He molecular beams and ultracold gases of 
  $^{133}$Cs, the discovery of an excited state 
  \cite{Lim1977a,Cornelius1986a,Esry1996a} remains an elusive goal. 
  Measurement of the $E_2$ three-body zero-energy resonance at about $-9\,$G 
  in Fig.~\ref{fig:133CsK3vsB} could provide an opportunity for such a 
  discovery. Ultracold gases of $^{133}$Cs prepared in the excited 
  $(F=3,m_F=-3)$ Zeeman state, however, are affected by inelastic atom loss 
  due to both two-body spin-relaxation \cite{Leo2000a,Chin2004a} and 
  three-body recombination. Such a coincidence of loss mechanisms could lead 
  to difficulties in the interpretation of measurements \cite{Roberts2000a}. 
 
  \begin{figure}[htb]
    \includegraphics[width=\columnwidth,clip]{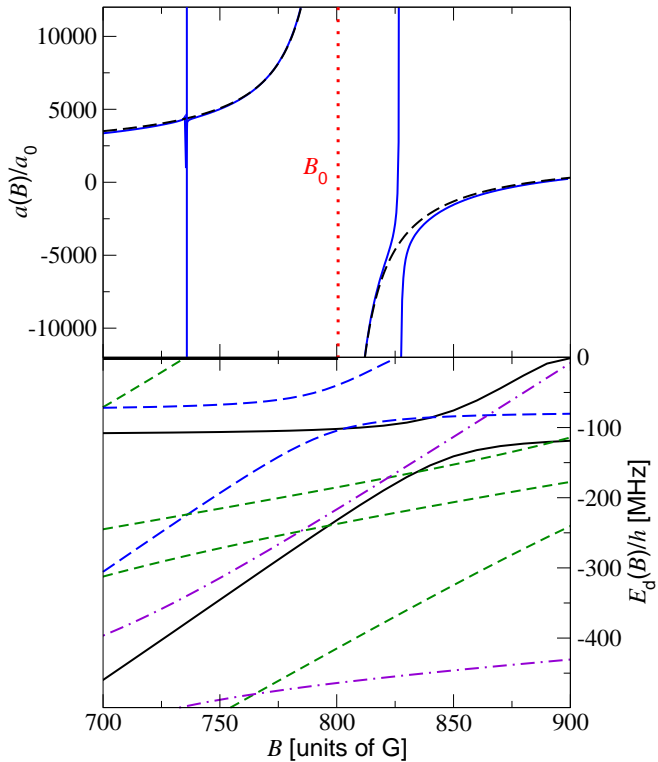}
    \caption{(Color online) Scattering length (upper panel) and diatomic 
      bound-state energies (lower panel) of $^{133}$Cs atoms 
      vs.~magnetic-field strength in the vicinity of 800\,G. The solid curve 
      in the upper panel refers to exact coupled-channels calculations 
      \cite{Chin2004a} including basis states of $s$- and $d$-wave symmetries, 
      while the dashed curve is a fit of Eq.~(\ref{aofB}) to the calculated 
      data. According to this fit, the position of the broadest singularity of 
      the scattering length is located at $B_0=800.6\,$G (vertical dotted 
      line). Comparatively narrow zero-energy resonances near 735\,G and 
      825\,G are due to dimer levels of predominantly $d$-wave symmetry shown 
      in the lower panel. Their energies with respect to the zero-energy 
      dissociation threshold are indicated by long-dashed, dashed and
      dot-dashed curves. These states are more weakly coupled to the continuum 
      of free atoms than $s$-wave dimer levels. For convenience, their mutual 
      coupling is neglected in these calculations. Solid curves in the lower 
      panel refer to those $s$-wave dimer levels that constitute the target 
      states included in our three-body recombination calculations. The 
      6$s$-dimer level \cite{Chin2004a} causing the broadest singularity of 
      the scattering length in the upper panel is unresolved on the energy 
      scale chosen in the figure.}
    \label{fig:133CsaofB800G}
  \end{figure}
  
  For this reason, we study three-body recombination in the vicinity of an as 
  yet unobserved broad diatomic zero-energy resonance of $^{133}$Cs gases 
  prepared in the $(F=3,m_F=+3)$ Zeeman ground state at comparatively high
  magnetic-field strengths on the order of 800\,G. Based on the 
  coupled-channels model of Ref.~\cite{Chin2004a} and a fit to 
  Eq.~(\ref{aofB}), we obtain an associated singularity of the scattering 
  length at $B_0=800.6\,$G, as illustrated in the upper panel of 
  Fig.~\ref{fig:133CsaofB800G}. This fit gives the background-scattering 
  length and resonance width to be $a_\mathrm{bg}=1918\,a_0$ and 
  $\Delta B=83.2\,$G, respectively, determining $a(B)$ via Eq.~(\ref{aofB}). 
  Comparatively narrow zero-energy resonances at about 735\,G and 825\,G
  in Fig.~\ref{fig:133CsaofB800G} are neglected in this treatment.
  
  The lower panel of Fig.~\ref{fig:133CsaofB800G} illustrates the complex 
  diatomic molecular energy spectrum of $^{133}$Cs in the vicinity of 800\,G. 
  Dashed curves are associated with states of predominantly $d$-wave 
  symmetry, which cause the two comparatively narrow zero-energy resonances 
  in the upper panel. Solid curves refer to the $s$-wave dimer levels, which 
  are most strongly coupled to the scattering continuum above the zero-energy 
  threshold. Among them the 6$s$-Fesh\-bach molecular dimer state of energy 
  $E_\mathrm{b}$ (not resolved in the lower panel of 
  Fig.~\ref{fig:133CsaofB800G}) causes the broad diatomic zero-energy 
  resonance whose calculated position is $800.6\,$G. In order to give 
  quantitative estimates for $K_3$, we use the parametric two-channel 
  representation of the $s$-wave dimer levels explained in the appendix, in 
  analogy to our treatment of three-body recombination at low magnetic fields. 
  
  \begin{figure}[htb]
    \includegraphics[width=\columnwidth,clip]{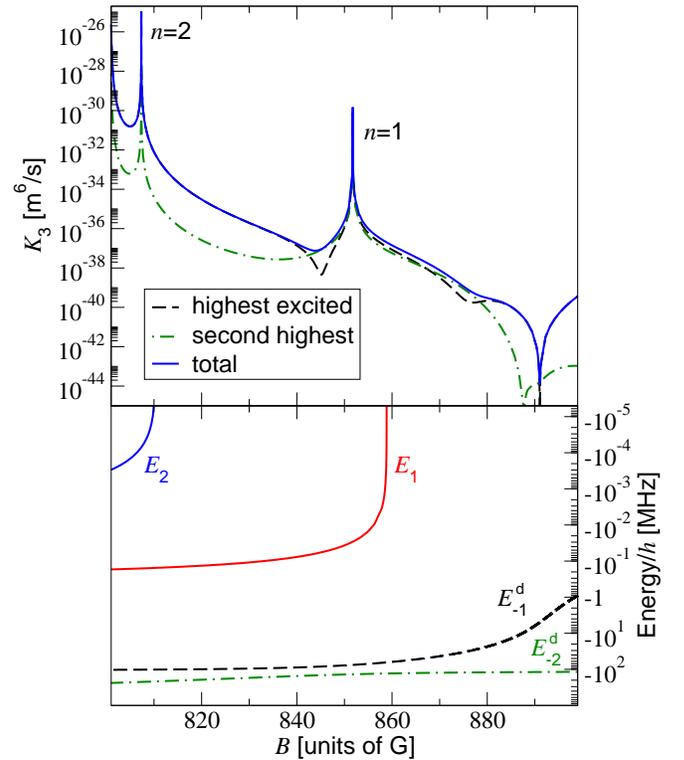}
    \caption{(Color online) 
    Calculated three-body recombination loss-rate constants $K_3$ (upper 
    panel) and energies of metastable trimer levels of $^{133}$Cs atoms
    labeled $E_1$ and $E_2$ (lower panel) vs.~the magnetic-field strength $B$
    on the high-field side of the 800.6\,G diatomic zero-energy resonance. For 
    each magnetic-field strength zero energy in the lower panel coincides with 
    the scattering threshold for three asymptotically free atoms. Similarly to 
    Fig.~\ref{fig:133CsK3vsB}, the dashed and dot-dashed curves labeled 
    $E^\mathrm{d}_{-1}$ and $E^\mathrm{d}_{-2}$ in the lower panel refer to 
    the $s$-wave dimer levels shown in the lower panel of
    Fig.~\ref{fig:133CsaofB800G}, which constitute the target states included 
    in our calculations of $K_3$. Dashed and dot-dashed curves in the upper 
    panel indicate their associated contributions to the total three-body 
    recombination loss-rate constant (solid curve). Three-body zero-energy 
    resonance peaks are labeled by the degree of excitation ($n=1,2$) of their 
    associated metastable Thomas-Efimov trimer levels of the lower panel.}
    \label{fig:133CsK3vsB800G}
  \end{figure}
  
  Figure~\ref{fig:133CsK3vsB800G} shows calculated three-body recombination 
  loss-rate constants as well as metastable Thomas-Efimov trimer levels 
  causing three-body zero-energy resonances in an ultracold cesium gas at 
  magnetic-field strengths $B$ in the vicinity of $800\,$G. Similarly to 
  Fig.~\ref{fig:133CsK3vsB}, these predictions refer to zero kinetic energy of 
  the three initially free colliding atoms. Based on such full calculations 
  (upper panel of Fig.~\ref{fig:133CsK3vsB800G}) as well as a simplifying 
  single-spin-channel treatment (lower panel of 
  Fig.~\ref{fig:133CsK3vsB800G}), we estimate the three-body zero-energy 
  resonance associated with the ground state ($n=1$) Thomas-Efimov level to 
  occur between 852\,G and 859\,G. Similarly to the results of 
  Fig.~\ref{fig:133CsK3vsB}, the deviations between the results of these two 
  completely different methods are less than 9\% of the diatomic resonance 
  width $\Delta B$ and mainly reflect the influence of the $E^\mathrm{d}_{-1}$ 
  and $E^\mathrm{d}_{-2}$ dimer levels (lower panel of 
  Fig.~\ref{fig:133CsK3vsB800G}) on the three-body energy spectrum. 
  Analogously, we estimate the position of the excited state ($n=2$) 
  Thomas-Efimov zero-energy resonance between 807\,G and 810\,G. While our 
  calculations do not account for the full complexity of the $^{133}$Cs 
  diatomic energy spectrum, as illustrated in the lower panel of 
  Fig.~\ref{fig:133CsaofB800G}, we believe the accuracy of these estimates to 
  be similar to those for the low magnetic-field strengths in 
  Fig.~\ref{fig:133CsK3vsB}. 
  
  \begin{figure}[htb]
    \includegraphics[width=\columnwidth,clip]{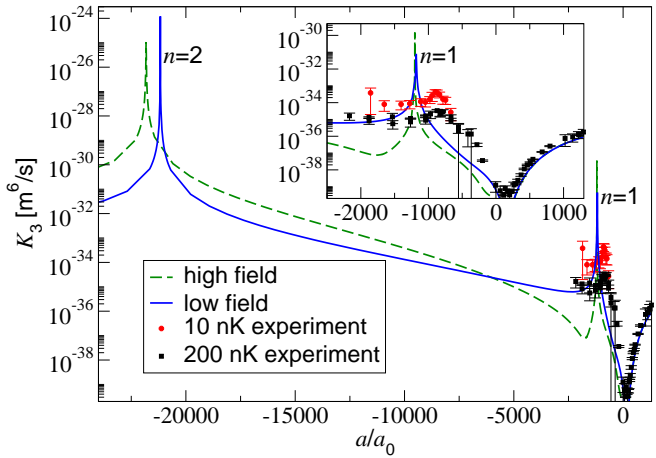}
    \caption{(Color online) 
      Comparison between the calculated total three-body recombination 
      loss-rate constants $K_3$ associated with the low-field (solid curve) 
      and high-field (dashed curve) diatomic zero-energy resonances of 
      $^{133}$Cs vs.~the scattering length, $a$. Filled circles and squares 
      refer to experimental data of Refs.~\cite{Kraemer2006a,Weber2003a} 
      observed at low magnetic-field strengths (see Fig.~\ref{fig:133CsK3vsB}) 
      in dilute gases of 10\,nK and 200\,nK, respectively. The range of 
      scattering lengths displayed covers the ground-state ($n=1$) and 
      excited-state ($n=2$) three-body zero-energy resonances of the 
      Thomas-Efimov spectrum. The inset shows an enlargement of the presently 
      experimentally accessed \cite{Kraemer2006a,Weber2003a} range of $a(B)$ 
      as well as the associated calculated and measured $K_3$ loss-rate 
      constants.}
    \label{fig:complowhigh}
  \end{figure}
  
  In Fig.~\ref{fig:complowhigh} we compare the total three-body recombination
  loss-rate constants, $K_3$, associated with the low-field and high-field 
  diatomic zero-energy resonances as functions of the scattering length, $a$. 
  The overall qualitative trends of the calculated $K_3$ near 800\,G (dashed 
  curve) tend to be similar to those at low fields (solid curve). In the 
  regime of negative scattering lengths, however, the functional dependences 
  on $a$ differ considerably between the two curves, demonstrating the strong 
  sensitivity of $K_3$ to the different $B$-dependent properties of the 
  associated, comparatively tightly bound diatomic target levels. Universal 
  behavior of $K_3$, which is insensitive to the $B$-dependent properties of  
  tightly bound diatomic target states, occurs only in the limit of large, 
  positive $a$, where both curves tend to agree. In this regime, three-body 
  recombination is strongly dominated by the weakly bound, highest excited, 
  vibrational dimer state, whose long-range wave function is determined mainly 
  by $a(B)$ in addition to the van der Waals dispersion coefficient, $C_6$ 
  \cite{Koehler2006a}. These parameters also determine the positions of 
  Thomas-Efimov three-body zero-energy resonances in our single-spin-channel 
  estimates of their associated  energy levels in the lower panels of 
  Figs.~\ref{fig:133CsK3vsB} and \ref{fig:133CsK3vsB800G} \cite{Stoll2005a}.

  We predict a pronounced minimum of $K_3$ to occur at about 891\,G in the 
  vicinity of zero scattering length. According to 
  Fig.~\ref{fig:133CsK3vsB800G}, a lower estimate of the associated loss-rate 
  constants is on the order of only $10^{-44}\,$m$^6$/s. Given the comparison 
  between theory and experiment of Fig.~\ref{fig:133CsK3vsB}, we believe that 
  recombination into dimer levels beyond those included in our calculations as 
  well as finite temperature effects could significantly increase the minimal 
  value of $K_3$. Similarly to the experiments of 
  Refs.~\cite{Cornish2000a,Weber2003b}, however, the associated minimal atom 
  loss may still allow Bose-Einstein condensation of $^{133}$Cs at high fields 
  using magnetically tunable diatomic interactions. This possibility would be 
  crucial to three-body recombination measurements at negative scattering 
  lengths using the technique reported in Ref.~\cite{Kraemer2006a}.
     
  \subsection{Thomas-Efimov spectrum of $^{85}$Rb}
   
  Both examples of diatomic zero-energy resonances of $^{133}$Cs discussed in 
  this paper were broad and entrance-channel dominated \cite{Koehler2006a}.
  Entrance-channel dominance allowed us to describe and interpret near 
  threshold three-atom resonance states using single-spin-channel models, 
  largely in accordance with Thomas' \cite{Thomas1935a} and Efimov's 
  \cite{Efimov1970a} original treatments. Their broad nature should be crucial 
  for a possible experimental resolution of several three-body zero-energy 
  resonances, given present limitations in magnetic-field control. In both of 
  these examples, however, the background-scattering length was positive. This 
  implies that the dominant three-body recombination peak refers to the 
  previously observed ground state of the Thomas-Efimov spectrum, whereas the 
  excited-state peaks occur at large negative scattering lengths 
  ($|a|>20000\,a_0$ in Fig.~\ref{fig:complowhigh}) and accordingly large $K_3$ 
  loss-rate constants. Consequently, fast depletion of an ultracold gas at 
  such near resonant magnetic-field strengths as well as thermal broadening of 
  peaks \cite{Incao2004a} could impose limitations on the feasibility of 
  associated measurements. Recent studies \cite{Incao2006a} therefore suggest 
  alternative measurements in mixtures of ultracold gases with unequal masses 
  of the constituent particles. In such systems the number of exotic 
  three-body energy levels depends sensitively on the associated mass ratios 
  (see Refs.~\cite{Amado1972a,Newton1982a} and references therein).
  
  \begin{figure}[htb]
    \includegraphics[width=\columnwidth,clip]{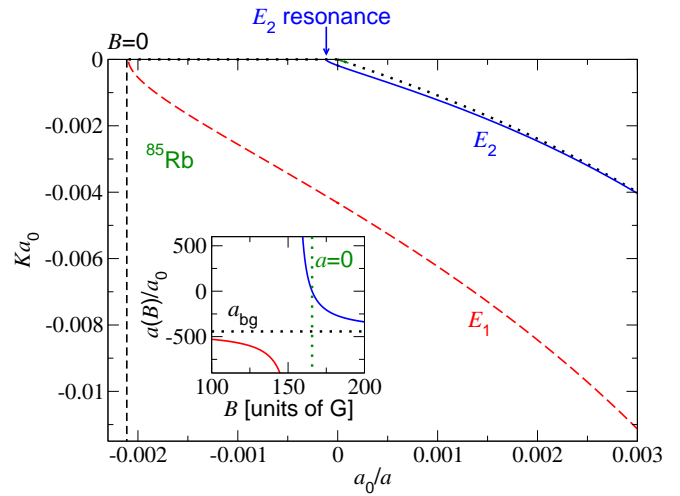}
    \caption{(Color online) 
    Efimov plot associated with $^{85}$Rb trimers \cite{Stoll2005a} in the 
    vicinity of the 155\,G diatomic zero-energy resonance. The solid and 
    dashed curves refer to the $E_2$ and $E_1$ levels, respectively, while the 
    dotted line indicates the energy of the Fesh\-bach molecular dimer state. 
    The inset shows the magnetic-field dependence of the scattering length 
    $a(B)$ for this atomic species with a negative background-scattering 
    length of $a_\mathrm{bg}=-443\,a_0$ \cite{Claussen2003a}.}
    \label{fig:Efimov85Rb}
  \end{figure}
  
  In accordance with Ref.~\cite{Stoll2005a}, ultracold gases with broad, 
  entrance-channel dominated diatomic zero-energy resonances 
  \cite{Koehler2006a} and negative background-scattering lengths may be best 
  suited for the observation of an excited trimer state in the original 
  context of Efimov's suggestion for identical Bose particles of equal mass 
  \cite{Efimov1970a}. An associated Efimov plot is shown in 
  Fig.~\ref{fig:Efimov85Rb} for the example of $^{85}$Rb in the vicinity of 
  155\,G. According to the inset, the inverse scattering length $1/a(B)$ 
  increases with increasing $B$ and extends from zero field (dashed vertical 
  line) to the zero of the scattering length at about 165.71\,G 
  \cite{Claussen2003a,Thompson2005a}. Consequently, a range of $1/a$ from 
  $1/a_\mathrm{bg}<0$ to $+\infty$ is experimentally accessible. As opposed to 
  the cesium case in the upper panel of Fig.~\ref{fig:compHeCs}, the ground 
  Thomas-Efimov level never reaches the threshold in the vicinity of 
  $B_0=155\,$G. Consequently, the only available three-body zero-energy 
  resonance refers to the excited and as yet unobserved $E_2$ level. 
  
  Besides the possibility of measuring an associated three-body recombination 
  loss-rate peak, the existence of the excited Thomas-Efimov state could be 
  verified by populating this trimer level using a magnetic-field sweep 
  technique \cite{Stoll2005a}. Such experiments might be performed in the gas 
  phase as well as in optical lattices with three atoms occupying each site 
  \cite{Jonsell2002a,Stoll2005a,Werner2006,Luu2006a}. Due to the larger 
  spatial extents of their bound-state wave functions \cite{Stoll2005a}, the 
  lifetimes of excited $E_2$ states of $^{85}$Rb$_3$, with respect to 
  spontaneous dissociation via either spin relaxation 
  \cite{Thompson2005a,Koehler2005a} or atom-dimer relaxation 
  \cite{Nielsen2002a,Jonsell2006a}, should be significantly longer than those 
  of ground states of the Thomas-Efimov spectrum. Such an increase in lifetime 
  of these ultracold trimer molecules might be sufficient for experimental 
  studies of their exotic properties.  

  \section{Conclusions}
  \label{sec:conclusions}
  
  We have shown by an intuitive argument as well as full numerical 
  calculations that the recent observation of a three-body zero-energy 
  resonance in ultracold $^{133}$Cs gases \cite{Kraemer2006a} refers to the 
  ground level of the Thomas-Efimov spectrum that has been discovered in 
  $^4$He molecular-beam experiments \cite{Schollkopf1994a}. Unlike $^4$He$_3$, 
  the metastable cesium trimer observed in the experiments of 
  Ref.~\cite{Kraemer2006a} is associated with a Borromean state 
  \cite{Zhukov1993a}, providing, to our knowledge, the first experimental 
  evidence for their existence in molecular physics. Our studies show that 
  resonance-enhanced three-body recombination in ultracold gases at negative 
  as well as small positive scattering lengths is sensitive to the strongly  
  magnetic-field dependent Zeeman-state compositions of the comparatively 
  tightly bound diatomic target states. This implies that fitting procedures 
  incorporating the magnetic-field dependence of the rate constants only 
  through the scattering length are incomplete and could therefore lead to 
  unreliable conclusions about universality of trimer levels. As opposed to 
  the magnitudes of the loss-rate constants, the mere magnetic-field positions 
  of Thomas-Efimov zero-energy resonances largely follow the qualitative 
  trends implied by three-body universality, irrespective of the details of 
  diatomic energy levels far away from the scattering threshold. 
  
  The long-standing problem of detecting excited Thomas-Efimov states 
  \cite{Cornelius1986a,Esry1996a,Bruhl2005a} remains an elusive goal. Based on 
  numerical predictions of three-body recombination loss-rate constants, we 
  have investigated the possibility of observing associated three-body 
  zero-energy resonances in ultracold $^{133}$Cs gases. Our studies focused on 
  a broad, entrance-channel dominated \cite{Koehler2006a} diatomic 
  zero-energy resonance, which is predicted to occur near 800\,G. In this 
  case, the background-scattering length is positive. We have shown that for 
  this reason an associated experiment would require observation of two 
  distinct three-body recombination loss-rate peaks. The excited state 
  three-body zero-energy resonance occurs at large negative scattering 
  lengths. Thermal broadening \cite{Incao2004a} as well as strong atom loss 
  could therefore impose limitations on the feasibility of associated 
  measurements. We reach the conclusion that ultracold gases involving 
  entrance-channel dominated diatomic zero-energy resonances with negative 
  background-scattering lengths, such as $^{85}$Rb, may be best suited for the 
  first observation of excited Thomas-Efimov trimers. Their detection might be 
  achieved by populating such metastable states, with sufficient lifetimes, 
  via magnetic-field sweeps \cite{Stoll2005a} in ultracold gases or in optical 
  lattices with three atoms occupying each site 
  \cite{Jonsell2002a,Stoll2005a,Werner2006,Luu2006a}.
  
  \section{Acknowledgments}
  
  We are grateful to Tobias Kraemer for providing measured data sets of 
  Refs.~\cite{Kraemer2006a,Weber2003a} to us, as well as to Vitaly Efimov for 
  his advice on signatures of universal three-body states in nature, 
  especially with regard to the three-nucleon problem. We thank Fran\c{c}oise 
  Masnou-Seeuws, Francesca Ferlaino, Hanns-Christoph N\"agerl, Rudolf Grimm, 
  Jos\'e D'Incao, Brett Esry, Chris Greene, Wieland Sch\"ollkopf, and Peter 
  Toennies for many helpful discussions. This research has been supported by 
  the Royal Society and the UK EPSRC.
  
  \appendix

  \section{AGS approach to three-body recombination}
  \label{app:AGS}
  
  \subsection{Scattering matrix}
  
  \subsubsection{Asymptotic scattering channels}
  
  Throughout this appendix, we consider collisions between three identical 
  Bose atoms. Three-body scattering is characterized by the arrangements of 
  the atoms on asymptotically large time scales before and after a collision. 
  These arrangements determine the asymptotic entrance- and exit-scattering 
  channels. Such channels can involve a bound molecular pair of atoms 
  asymptotically spatially separated from the third atom or three free atoms. 
  In the case of three-body recombination the entrance channel consists of
  three asymptotically free atoms, whereas the exit channel contains a 
  diatomic molecule and a free atom. 

  \begin{figure}[htbp]
    \includegraphics[width=\columnwidth,clip]{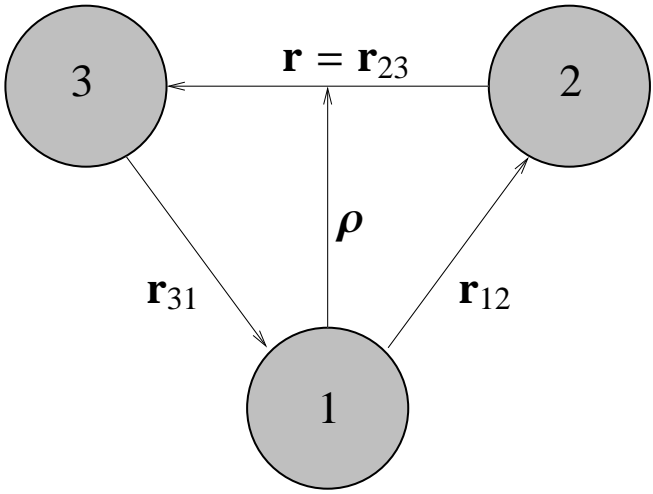}
    \caption{Jacobi coordinates of the relative motion of three atoms. The set 
      of coordinates $\boldsymbol{\rho}=\boldsymbol{\rho}_1$ and 
      $\mathbf{r}=\mathbf{r}_1$ is selected in such a way that it describes 
      the hypothetical situation of an interacting pair of atoms $(2,3)$, 
      whereas atom 1 is considered as a spectator. Two other sets of these 
      coordinates describing different arrangements of the atoms can be 
      obtained through cyclic permutations of the atomic indices.}
    \label{fig:Jacobi}
  \end{figure}
  
  To characterize the positions of the atoms in the different asymptotic 
  scattering channels, we choose the Jacobi coordinates of 
  Fig.~\ref{fig:Jacobi} in the center-of-mass frame. These coordinates are 
  particularly well suited to describe asymptotic arrangements involving a 
  bound pair and a free atom. In the following, we label the sets of 
  coordinates by the Greek index $\alpha=1,2,3$ of the free atom. For 
  convenience, the particular set of coordinates shown in 
  Fig.~\ref{fig:Jacobi} will be denoted, without explicit reference to its 
  index, simply by $(\boldsymbol{\rho},\mathbf{r})$. 

  \subsubsection{Transition matrices}

  We assume the three atoms to interact pairwise depending on their internal 
  Zeeman states \cite{Smirne2006a}. Consequently, the three-body Hamiltonian 
  in the barycentric frame, 
  \begin{equation}
    H=H_0+V_1+V_2+V_3,
    \label{H3B} 
  \end{equation}
  is comprised of a kinetic energy contribution, 
  \begin{equation}
    H_0=|\mathrm{bg}\rangle
    \left[
    -\frac{\hbar^2}{2(\frac{2}{3} m)}\nabla_{\boldsymbol{\rho}}^2
    -\frac{\hbar^2}{2(\frac{m}{2})}\nabla_{\mathbf{r}}^2
    \right]
    \langle\mathrm{bg}|,
  \end{equation}
  in addition to the pair potentials $V_1$, $V_2$, and $V_3$. Here $m$ is the 
  atomic mass, $|\mathrm{bg}\rangle$ denotes the entrance-channel spin product 
  state, consisting of the three one-body Zeeman states in which the atoms of 
  the gas are prepared, and $V_1$ is the potential of the atom pair $(2,3)$. 
  The remaining interactions $V_2$ and $V_3$ are determined through cyclic 
  permutations of the atomic indices. Each potential couples entrance- and 
  closed-channel Zeeman state configurations in Fesh\-bach resonance-enhanced 
  collisions \cite{Smirne2006a}. As the entire physics described in this paper
  involves energy ranges in the close vicinity of the scattering threshold, 
  and accordingly large interparticle distances, we have neglected 
  short-ranged, genuinely three-body forces in our approach 
  \cite{Esry1996a,Stoof1989a}.  

  The probabilities for transitions between the different asymptotic 
  scattering channels can be conveniently described in terms of the scattering 
  matrix $S$. To determine the $S$-matrix, we follow the approach by Alt, 
  Grassberger and Sandhas (AGS) \cite{Alt1967a,Sandhas1972a} and introduce the 
  following Green's functions associated with the different asymptotic 
  arrangements of the atoms:
  \begin{align}
    G_0(z)&=(z-H_0)^{-1},\\
    G_\alpha(z)&=(z-H_0-V_\alpha)^{-1}.
  \end{align}
  Here $z$ is a complex variable with the dimension of an energy. Throughout 
  this appendix this variable is given by a limit, $z=E_\mathrm{i}+i0$, 
  indicating that the energy of the three initially noninteracting atoms, 
  $E_\mathrm{i}$, is approached from the upper half of the complex plane. The 
  free Green's function $G_0(z)$ is associated with three asymptotically free 
  atoms, while $G_\alpha(z)$ describes the configuration in which the atom 
  with the index $\alpha$ is free and the remaining atom pair interacts via 
  the potential $V_\alpha$. The complete Green's function 
  \begin{equation}
    G(z)=(z-H)^{-1}
  \end{equation}
  implicitly determines the AGS transition matrix $U_{\alpha\beta}(z)$ through 
  the relation:
  \begin{equation}
    G(z)=\delta_{\alpha\beta}G_\beta(z)+
    G_\alpha(z)U_{\alpha\beta}(z)G_\beta(z).
    \label{defUalphabeta}
  \end{equation}
  Here the indices $\beta=0$ and $\alpha=0$ are associated with entrance and 
  exit channels involving three free atoms, respectively. The indices 
  $\beta,\alpha=1,2,3$ correspond to arrangements in which the atom with the
  index $\beta$ or $\alpha$ is free, respectively, while the remaining atom 
  pair is bound to a molecule. 
  
  The transition matrices $U_{\alpha\beta}(z)$ determine the elements of the 
  $S$-matrix associated with the different asymptotic scattering channels. 
  The general derivations of Refs.~\cite{Alt1967a,Sandhas1972a} show, in 
  particular, that the $S$-matrix element for three-body recombination into a 
  diatomic molecular bound state $|\phi_\mathrm{d}\rangle$ is given by the 
  formula:
  \begin{align}
    \nonumber
    _\alpha\langle \mathbf{q}_\mathrm{f},\phi_\mathrm{d}|
    S|\mathbf{q}_\mathrm{i},\mathbf{p}_\mathrm{i},\mathrm{bg}\rangle
    =&-2\pi i \delta(E_\mathrm{f}-E_\mathrm{i})\\
    &\times {_\alpha\langle} \mathbf{q}_\mathrm{f},\phi_\mathrm{d}|
    U_{\alpha 0}(z)
    |\mathbf{q}_\mathrm{i},\mathbf{p}_\mathrm{i},\mathrm{bg}\rangle.
    \label{Smatrix}
  \end{align}
  Here $|\mathbf{q}_\mathrm{i},\mathbf{p}_\mathrm{i},\mathrm{bg}\rangle=
  |\mathbf{q}_\mathrm{i}\rangle|\mathbf{p}_\mathrm{i}\rangle
  |\mathrm{bg}\rangle$ and $|\mathbf{q}_\mathrm{f},\phi_\mathrm{d}
  \rangle_\alpha=|\mathbf{q}_\mathrm{f}\rangle_\alpha|\phi_\mathrm{d}
  \rangle_\alpha$ are the general unsymmetrized initial and final 
  momentum-normalized energy states associated with the entrance and exit 
  channels, respectively. Consequently, $\mathbf{p}$ and $\mathbf{q}$ denote 
  the momenta associated with the Jacobi coordinates $\mathbf{r}$ and 
  $\boldsymbol{\rho}$, respectively. This implies, for instance, that 
  $\mathbf{q}_\mathrm{f}$ is the momentum of the center of mass of the 
  molecule produced relative to the free atom with index $\alpha$, and 
  $\langle\boldsymbol{\rho}|\mathbf{q}_\mathrm{f}\rangle=
  \exp(i\,\mathbf{q}_\mathrm{f}\cdot\boldsymbol{\rho}/\hbar)/
  (2\pi\hbar)^{3/2}$ is the associated plane wave. Similarly, 
  $|\phi_\mathrm{d}\rangle_\alpha$ is the diatomic molecular state of the 
  interacting atom pair in the exit channel. We have arbitrarily chosen the 
  Jacobi momenta describing the entrance-channel state in accordance with the 
  selection of atom 1 in Fig.~\ref{fig:Jacobi} and omitted the index 
  associated with this particular arrangement of particles. The asymptotic 
  energies of the three atoms in the barycentric frame are symmetric with 
  respect to permutations of particles and related to the Jacobi momenta by 
  the formulae:
  \begin{align}
    \label{Ei}
    E_\mathrm{i}&=\frac{3q_\mathrm{i}^2}{4m}+
    \frac{p_\mathrm{i}^2}{m},\\
    E_\mathrm{f}&=\frac{3q_\mathrm{f}^2}{4m}+E_\mathrm{d}.
    \label{Ef}
  \end{align}
  Here $E_\mathrm{d}$ is the binding energy associated with the diatomic 
  molecular target state $|\phi_\mathrm{d}\rangle_\alpha$. Consequently, the 
  entrance- and exit-channel states solve the Schr\"odinger equations 
  $H_0|\mathbf{q}_\mathrm{i},\mathbf{p}_\mathrm{i},\mathrm{bg}\rangle=
  E_\mathrm{i}|\mathbf{q}_\mathrm{i},\mathbf{p}_\mathrm{i},\mathrm{bg}\rangle$
  and $(H_0+V_\alpha)|\mathbf{q}_\mathrm{f},\phi_\mathrm{d}\rangle_\alpha=
  E_\mathrm{f}|\mathbf{q}_\mathrm{f},\phi_\mathrm{d}\rangle_\alpha$, 
  respectively. The $S$-matrix elements for transitions between other 
  scattering channels can be represented in terms of the transition matrices 
  of Eq.~(\ref{defUalphabeta}) through formulae similar to Eq.~(\ref{Smatrix}) 
  \cite{Alt1967a,Sandhas1972a}. 
  
  \subsection{Three-body recombination loss-rate constant}
  
  \subsubsection{Probability for the recombination of three atoms}
  
  To determine the three-body recombination loss-rate constant, we consider 
  three initially free atoms in a large box of volume $\mathcal{V}$ which will 
  later be taken to be infinite. The probability $P_\mathrm{fi}$ for 
  recombination into the molecular bound state $|\phi_\mathrm{d}\rangle$, in a 
  time interval of length $\Delta t$, can be obtained using Fermi's golden 
  rule. This yields:
  \begin{align}
    \nonumber
    P_\mathrm{fi}=&\frac{18}{\hbar^2}
    \frac{(2\pi\hbar)^6}{\mathcal{V}^2}
    \int d\mathbf{q}_\mathrm{f} 
    \left(
    \frac{
      \sin\left[\frac{1}{2}(E_\mathrm{f}-E_\mathrm{i})\Delta t/\hbar\right]}
	 {\frac{1}{2}(E_\mathrm{f}-E_\mathrm{i})/\hbar}
	 \right)^2\\
	 &\times
	 \left|
	 {_1\langle} 
	 \mathbf{q}_\mathrm{f},\phi_\mathrm{d}|U_{1,0}(z)
	 \mathcal{S}|\mathbf{q}_\mathrm{i},\mathbf{p}_\mathrm{i},
	 \mathrm{bg}
	 \rangle
	 \right|^2.  
	 \label{Pfi}
  \end{align}
  We note that the complex argument ``$z=E_\mathrm{i}+i0$'' of the transition 
  matrix $U_{1,0}(z)$ indicates that the physical energy $E_\mathrm{i}$ is 
  approached from the upper half of the complex plane. To account for the 
  identical nature of the Bose atoms, the entrance-channel state 
  $|\mathbf{q}_\mathrm{i},\mathbf{p}_\mathrm{i},\mathrm{bg}\rangle_1 
  (2\pi\hbar)^3/\mathcal{V}$ in Eq.~(\ref{Pfi}) is symmetrized by the
  projection 
  \begin{equation}
    \mathcal{S}=\frac{1}{6}\sum_{\mathcal{P}\in \mathcal{S}_3}\mathcal{P},
  \end{equation}
  where the sum extends over all $3!=6$ permutations $\mathcal{P}$ of the 
  atoms. Due to the symmetric nature of the projected initial wave function in 
  Eq.~(\ref{Pfi}) the index $\alpha=1$ associated with the final atomic 
  arrangement is arbitrary.   

  The approximations underlying the Fermi golden rule argument and the 
  associated continuum limit of the energy states of the box determine the 
  range of validity of Eq.~(\ref{Pfi}). The time scale $\Delta t$ is limited 
  from above by the requirement that phenomena associated with secondary 
  collisions of atoms due to reflection from the boundaries of the box are 
  negligible. This implies that $\Delta t$ must be much smaller than the 
  ground-state period of the box. This gives the restriction
  \begin{equation}
    \Delta t\ll \hbar/E_\mathrm{g}^\mathrm{box}=
    2m\mathcal{V}^{2/3}/(\pi^2\hbar),
    \label{upperboundDeltat}
  \end{equation}
  where $E_\mathrm{g}^\mathrm{box}=\hbar^2\pi^2/(2m\mathcal{V}^{2/3})$ is the 
  single-particle ground-state energy of the box. Given that ultracold 
  collision energies $E_\mathrm{i}$ are usually far below the breakup 
  threshold, i.e.~$E_\mathrm{i}\ll|E_\mathrm{d}|$, a simple calculation using 
  Eqs.~(\ref{Ei}) and (\ref{Ef}) shows that the condition 
  \begin{equation}
    \hbar/\left|E_\mathrm{d}\right|\ll\Delta t
    \label{lowerboundDeltat}
  \end{equation}
  allows us to perform the formal limit 
  \begin{equation}
    \left(
    \frac{
      \sin\left[\frac{1}{2}(E_\mathrm{f}-E_\mathrm{i})\Delta t/\hbar\right]}
	 {\frac{1}{2}(E_\mathrm{f}-E_\mathrm{i})/\hbar}
	 \right)^2
	 \underset{\Delta t\to\infty}{\sim}2\pi\hbar
	 \delta(E_\mathrm{f}-E_\mathrm{i})
	 \Delta t
	 \label{energyconservation}
  \end{equation}
  in the integral over the final momentum $\mathbf{q}_\mathrm{f}$ of 
  Eq.~(\ref{Pfi}). The applicability of Eq.~(\ref{energyconservation}) ensures 
  that the transition probability per unit time, 
  i.e.~$P_\mathrm{fi}/\Delta t$, is independent of $\Delta t$ within the 
  limitations set by Eqs.~(\ref{upperboundDeltat}) and 
  (\ref{lowerboundDeltat}). We note that the lower bound on $\Delta t$ set by 
  Eq.~(\ref{lowerboundDeltat}) is also necessary for the validity of 
  Eq.~(\ref{Pfi}). The time scale $\hbar/\left|E_\mathrm{d}\right|$ is 
  therefore characteristic for a three-body recombination event.

  \subsubsection{Three-body recombination loss rate for a thermal Bose gas}
  
  The interpretation of atom loss observed in a dilute atomic gas in terms of 
  rate phenomena associated with any chemical reaction crucially relies upon 
  the possibility of introducing coarse-grained scales in space and time. To 
  this end, we divide the volume of the gas into regions of virtually constant 
  density $n(\mathbf{x})$ characterized by their central position 
  $\mathbf{x}$. The volume $\mathcal{V}$ of each region sets the scale for the 
  unit of length in the coarse-graining approach. The unit of time $\Delta t$ 
  is limited by Eqs.~(\ref{upperboundDeltat}) and (\ref{lowerboundDeltat}) 
  except that the volume of the box in Eq.~(\ref{upperboundDeltat}) needs to 
  be replaced by $[n(\mathbf{x})]^{-1}$. This replacement rules out collisions 
  of three specific colliding atoms with any other constituents of the gas 
  during a time interval of length $\Delta t$. 
  Equations~(\ref{upperboundDeltat}) and (\ref{lowerboundDeltat}) thus 
  determine the condition for the applicability of a rate treatment of 
  three-body recombination in a dilute gas to be:
  \begin{equation}
    |E_\mathrm{d}|\gg [n(\mathbf{x})]^{2/3}\hbar^2/m.
    \label{conditionrate}
  \end{equation}  
  In the absence of diatomic zero-energy resonances Eq.~(\ref{conditionrate}) 
  is fulfilled for most realistic sets of physical parameters of dilute 
  ultracold gases. Magnetic tuning of the bound-state energy $E_\mathrm{b}$ of 
  the highest excited vibrational state, however, may violate this condition. 

  Given that Eq.~(\ref{conditionrate}) applies, we follow the treatment of 
  Ref.~\cite{Moerdijk1996a} to derive the loss-rate constant for three-body 
  recombination. This treatment relies upon a summation of the microscopic 
  transition probabilities of Eq.~(\ref{Pfi}), per unit time $\Delta t$, over 
  the number $\left({N}\atop{3}\right)\approx N^3/6$ of atomic triplets in a 
  region with $N$ atoms. This yields the number of recombination events. Under 
  the assumption that all three atoms are lost in each recombination event, 
  the associated loss-rate constant in a thermal ultracold Bose gas is given 
  by the formula:
  \begin{align}
    \nonumber
  K_3&=\frac{1}{2}\mathcal{V}^2 P_\mathrm{fi}/\Delta t\\
  &=\frac{12\pi m}{\hbar} (2\pi\hbar)^6 q_\mathrm{f}
  \int d\Omega_\mathrm{f}
  \left|
  {_1\langle} \mathbf{q}_\mathrm{f},\phi_\mathrm{d}|U_{1,0}(z)
  |0\rangle\right|^2.
  \label{K3thermal}
  \end{align}
  Here $d\Omega_\mathrm{f}$ denotes the angular component of 
  $d\mathbf{q}_\mathrm{f}$. In the derivation of Eq.~(\ref{K3thermal}) we have 
  taken advantage of the fact that the transition amplitude  
  ${_1\langle}\mathbf{q}_\mathrm{f},\phi_\mathrm{d}|U_{1,0}(E_\mathrm{i}+i0)
  \mathcal{S}|\mathbf{q}_\mathrm{i},\mathbf{p}_\mathrm{i},\mathrm{bg}\rangle$ 
  of Eq.~(\ref{Pfi}) is often insensitive with respect to typical collision 
  energies $E_\mathrm{i}$ of a cold gas. For convenience, we have thus 
  approximated the entrance-channel state by the zero-momentum plane wave 
  $|0\rangle$ of the relative motion of three noninteracting entrance-channel 
  atoms and performed the limit $z=i0$. The energy conservation implied by 
  Eq.~(\ref{energyconservation}) then determines the modulus of the final 
  momentum of the diatomic molecule relative to the third, free, atom to be:
  \begin{equation}
    q_\mathrm{f}=2\sqrt{m\left|E_\mathrm{d}\right|/3}.
    \label{qf}
  \end{equation}
  The rate equation for the total number of atoms in a cold gas, $N(t)$, can 
  be obtained through averaging the local rate equations over all regions of 
  constant density. This local density approach yields 
  Eq.~(\ref{rateequation}). If recombination events occur in a Bose-Einstein 
  condensate then the three-body recombination rate $K_3\langle n^2(t)\rangle$ 
  needs to be divided by a factor of $3!=6$ \cite{Kagan1985a,Burt1997a}, 
  because the zero-energy entrance-channel state $|0\rangle$ is already 
  symmetric with respect to permutations of atoms \cite{Moerdijk1996a}. 

  \subsection{AGS method}
  
  \subsubsection{AGS equations for transition matrices}
  
  In the following, we describe the general approach we have employed to 
  determine the transition amplitude of Eq.~(\ref{K3thermal}) based on the AGS 
  method \cite{Alt1967a,Sandhas1972a}. Other techniques for solving the 
  three-body Schr\"odinger equation in the context of ultracold collisions
  often rely upon hyperspherical-coordinate representations (see, e.g., 
  Refs.~\cite{Nielsen1999a,Esry1999a,Suno2002a,Soldan2002a,Blandon2006a}). 
  According to Refs.~\cite{Alt1967a,Sandhas1972a}, all complete transition 
  matrices fulfill the coupled AGS equations:
  \begin{align}
    \label{AGSprior}
    U_{\alpha\beta}(z)&=(1-\delta_{\alpha\beta})G_0^{-1}(z)+
    \sum_{{\gamma=1}\atop{\gamma\neq\beta}}^{3}
    U_{\alpha\gamma}(z)G_\gamma(z)V_\gamma,\\
    U_{\alpha\beta}(z)&=(1-\delta_{\alpha\beta})G_0^{-1}(z)+
    \sum_{{\gamma=1}\atop{\gamma\neq\alpha}}^{3}
    V_\gamma G_\gamma(z)U_{\gamma\beta}(z).
    \label{AGSpost}
  \end{align}
  Here both Eq.~(\ref{AGSprior}) and Eq.~(\ref{AGSpost}) determine the 
  transition matrices $U_{\alpha\beta}(z)$ with indices $\alpha,\beta=1,2,3$. 
  We shall refer to them as the prior and post versions of the AGS equations, 
  respectively. Given the solution to these equations we insert $\beta=0$ into 
  the prior version of Eq.~(\ref{AGSprior}). This yields:
  \begin{equation}
    U_{\alpha 0}(z)=G_0^{-1}(z)+\sum_{\gamma=1}^{3}U_{\alpha\gamma}(z)
    G_\gamma(z)V_\gamma.
    \label{AGSrecombination}
  \end{equation}
  Taking advantage of the invariance of the zero-energy plane-wave state with 
  respect to permutations of the atoms, the AGS equation 
  (\ref{AGSrecombination}) gives the transition amplitude of
  Eq.~(\ref{K3thermal}) to be:
  \begin{align}
    \nonumber
    {_1\langle} \mathbf{q}_\mathrm{f},\phi_\mathrm{d}|U_{1,0}(z)
    |0\rangle
    =&\sum_{\beta=1}^{3}
    {_1\langle} \mathbf{q}_\mathrm{f},\phi_\mathrm{d}|
    U_{1\beta}(z)G_\beta(z)V_\beta
    |0\rangle\\
    =&\sum_{\alpha=1}^{3}
    {_\alpha\langle} \mathbf{q}_\mathrm{f},\phi_\mathrm{d}|
    U_{\alpha 1}(z)G_1(z)V_1
    |0\rangle.
    \label{AGSamplitude}
  \end{align}
  We note that each transition matrix $U_{\alpha\beta}(z)$ is equivalent to 
  the complete three-body Green's function due to Eq.~(\ref{defUalphabeta}). 
  The solutions of the coupled sets of equations (\ref{AGSprior}) as well as 
  (\ref{AGSpost}) thus determine all energy levels of the three-body 
  Hamiltonian rather than just the transition amplitudes between asymptotic 
  channel states.

  \subsubsection{AGS equations for the three-body recombination 
    transition amplitude}
  
  To render the problem of determining the exact amplitude of 
  Eq.~(\ref{AGSamplitude}) into a more practical form, we introduce the 
  following wave functions:
  \begin{align}
    |\psi_{\alpha 1}\rangle&=U_{\alpha 1}(z)G_1(z)V_1
    |0\rangle,\\
    |\psi_{\alpha 1}^\mathrm{i}\rangle&=G_0^{-1}(z)G_1(z)V_1
    |0\rangle 
    (1-\delta_{\alpha 1}).
  \end{align}
  Here the index $\alpha$ assumes the values $\alpha=1,2,3$. A straightforward 
  calculation using Eq.~(\ref{AGSpost}) shows that $|\psi_{\alpha 1}\rangle$ 
  is determined by the coupled set of Faddeev-type \cite{Faddeev1961a} 
  equations:
  \begin{equation}
    |\psi_{\alpha 1}\rangle=|\psi_{\alpha 1}^\mathrm{i}\rangle
    +\sum_{{\gamma=1}\atop{\gamma\neq\alpha}}^{3}
    V_\gamma G_\gamma(z)
    |\psi_{\gamma 1}\rangle.
    \label{AGSpsialpha1}
  \end{equation}
  The solution of Eq.~(\ref{AGSpsialpha1}) gives the transition amplitude for 
  three-body recombination to be:
  \begin{equation}
    {_1\langle} \mathbf{q}_\mathrm{f},\phi_\mathrm{d}|U_{1,0}(z)
    |0\rangle
    =\sum_{\alpha=1}^{3}
    {_\alpha\langle} \mathbf{q}_\mathrm{f},\phi_\mathrm{d}|
    \psi_{\alpha 1}\rangle.
    \label{AGSamplitudepsialpha1}
  \end{equation}
  We use Eq.~(\ref{AGSpsialpha1}) to determine the associated loss-rate 
  constant of Eq.~(\ref{K3thermal}), the implementation of which we discuss
  in Subsection~\ref{subsec:implementation}. In the following subsection, we 
  first provide a detailed description of the interatomic potentials.
 
  \subsection{Resonance-enhanced three-body recombination}
  
  \subsubsection{Three-body spin channels}
  
  Ultracold resonance-enhanced diatomic interactions 
  \cite{Stwalley1976a,Tiesinga1993a,Koehler2006a} usually occur due to 
  coupling between the entrance-spin-channel and a single energetically closed 
  channel. For all the broad diatomic zero-energy resonances described in this 
  paper the physical origin of this coupling is predominantly spin exchange 
  \cite{Stoof1988a,Tiesinga1993a}. In the context of three-body collisions the 
  resonant closed channel is characterized by a Zeeman-state configuration, 
  $|\mathrm{cl}\rangle_\alpha$. Here the index $\alpha=1,2,3$ indicates that 
  atom $\alpha$ is in the Zeeman state in which the Bose gas is prepared, 
  whereas the remnant atom pair is in the closed-channel spin state, whose 
  strong coupling to the entrance channel causes a diatomic zero-energy 
  resonance \cite{Stwalley1976a,Tiesinga1993a,Koehler2006a}. The three-body 
  spin state is orthogonal to the product state $|\mathrm{bg}\rangle$, i.e. 
  ${_\alpha\langle}\mathrm{cl}|\mathrm{bg}\rangle=0$. In the case of, e.g., 
  the $^{133}$Cs low-field zero-energy resonance of Figs.~\ref{fig:133CsEbvsB} 
  and \ref{fig:133CsK3vsB} the resonant diatomic closed channel is comprised 
  of Zeeman states mainly from the upper $F=4$ hyperfine level, whereas the 
  atoms are prepared in the $(F=3,m_F=3)$ Zeeman ground state. This implies 
  the orthogonality relation,  
  \begin{equation}
    {_\alpha\langle}\mathrm{cl}|\mathrm{cl}\rangle_\beta=\delta_{\alpha\beta},
    \label{orthogonalitycl}
  \end{equation}
  for all indices $\alpha,\beta=1,2,3$ of the three-body closed-channel spin 
  states. We assume in the following that the resonant interchannel coupling 
  fulfills Eq.~(\ref{orthogonalitycl}), i.e. no single-atom Zeeman state is 
  shared between the diatomic entrance and closed channels. This assumption 
  does not affect our description of diatomic energy levels for which the 
  closed channel is always orthogonal to the entrance-spin-channel, even when 
  single-particle Zeeman states are shared between them. It facilitates, 
  however, our calculations of three-body recombination loss-rate constants in 
  $^{133}$Cs gases and is strictly valid only for the low-field resonance 
  shown in Figs.~\ref{fig:133CsEbvsB} and \ref{fig:133CsK3vsB}. 
  
  \subsubsection{Single-resonance approach}
  
  \begin{figure}[htb]
    \includegraphics[width=\columnwidth,clip]{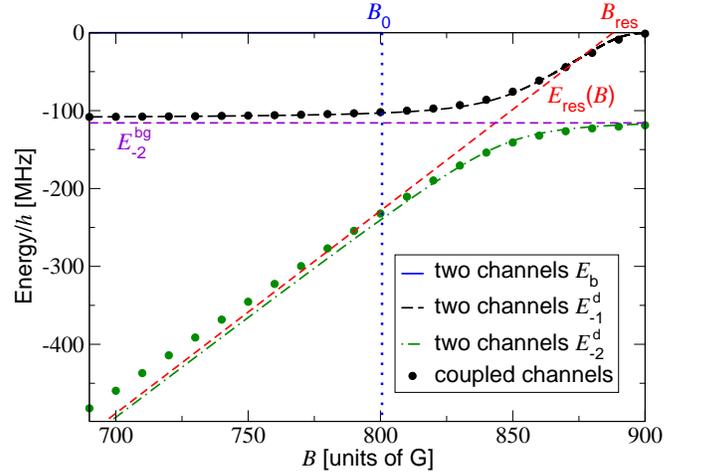}
    \caption{(Color online) Comparison between coupled-channels (filled 
      circles) and two-channel single-resonance (long-dashed and dot-dashed 
      curves) approaches to the highest excited vibrational $s$-wave states of 
      $^{133}$Cs$_2$ vs.~the magnetic-field strength $B$ in the vicinity of 
      800\,G. The dashed line labeled $E_\mathrm{res}(B)$ indicates the bare 
      Fesh\-bach resonance-state energy, while the horizontal dashed line 
      refers to the second highest excited bare vibrational level of the 
      background-scattering potential, $E_{-2}^\mathrm{bg}$. The energies of 
      the associated highest excited bare level, $E_{-1}^\mathrm{bg}$, and of 
      the Fesh\-bach molecular dimer state, $E_\mathrm{b}(B)$, are not 
      resolved. The point of degeneracy of $E_\mathrm{b}$ with the scattering 
      threshold (zero energy) is indicated by the vertical dotted line labeled 
      $B_0$. It coincides with the position of the broad singularity of the 
      scattering length in Fig.~\ref{fig:133CsaofB800G}. This measurable 
      zero-energy resonance position is distinct from the zero-energy crossing 
      point of the bare Fesh\-bach-resonance level, $B_\mathrm{res}$. The 
      deviation at lower $B$-fields between the coupled-channels and 
      two-channel $E_{-2}^\mathrm{d}$ levels is mainly due to a strong avoided 
      crossing of dressed, coupled-channels $s$-states in the vicinity of 
      600\,G, which is not included in the two-channel approach.}
    \label{fig:EbvsBccvs2c800G}
  \end{figure}
  
  Resonant enhancement of ultracold collisions is often well described in 
  terms of near degeneracy of a single diatomic closed-channel energy state, 
  $|\phi_\mathrm{res}\rangle$, with the dissociation threshold of the 
  entrance channel \cite{Stwalley1976a,Tiesinga1993a,Koehler2006a}. In an 
  extension of this concept to three-body scattering the resonance state, 
  $|\phi_\mathrm{res},\mathrm{cl}\rangle_\alpha$, refers to the situation in 
  which a pair of atoms occupies the state $|\phi_\mathrm{res}\rangle$, 
  whereas the third atom with index $\alpha$ plays the role of a spectator. 
  Throughout this appendix we choose $|\phi_\mathrm{res}\rangle$ to be unit 
  normalized, i.e.~$\langle\phi_\mathrm{res}|\phi_\mathrm{res}\rangle=1$. In 
  accordance with such a single-resonance approach, the pair potentials of the 
  three-body Hamiltonian~(\ref{H3B}) are of the following general form 
  \cite{Smirne2006a}:
  \begin{align}
    \nonumber
    V_\alpha=&|\mathrm{bg}\rangle V_\alpha^\mathrm{bg} 
    \langle\mathrm{bg}|
    +|\phi_\mathrm{res},\mathrm{cl}\rangle_\alpha\, E_\mathrm{res}(B)\, 
    {_\alpha}\langle\phi_\mathrm{res},\mathrm{cl}|\\
    \nonumber
    &+W_\alpha |\phi_\mathrm{res},\mathrm{bg}\rangle_\alpha\, 
    {_\alpha}\langle\phi_\mathrm{res},\mathrm{cl}|\\
    &+|\phi_\mathrm{res},\mathrm{cl}\rangle_\alpha\,
    {_\alpha}\langle\phi_\mathrm{res},\mathrm{bg}|W_\alpha.
    \label{potentials}	
  \end{align}
  Here $V_\alpha^\mathrm{bg}$ is the background-scattering potential, 
  $W_\alpha(\mathbf{r}_\alpha)$ is the potential energy of interchannel 
  coupling depending on the relative position $\mathbf{r}_\alpha$ of atoms of 
  an interacting pair, and $E_\mathrm{res}(B)$ is the resonance-state energy 
  \cite{Koehler2006a}. As we have chosen zero energy to coincide with the 
  scattering threshold for three asymptotically free atoms, the three-body 
  Hamiltonian (\ref{H3B}) depends on the magnetic-field strength only through 
  $E_\mathrm{res}(B)$. To a good approximation this dependence is determined 
  by \cite{Moerdijk1995a}:
  \begin{equation}
    E_\mathrm{res}(B)=\mu_\mathrm{res}(B-B_\mathrm{res}).
    \label{EresofB}
  \end{equation}
  Here $\mu_\mathrm{res}$ is the difference in magnetic moment of the 
  resonance state and a pair of asymptotically separated entrance-channel 
  atoms, and $B_\mathrm{res}$ is the crossing point between $E_\mathrm{res}$ 
  and the entrance-channel scattering threshold, 
  i.e.~$E_\mathrm{res}(B_\mathrm{res})=0$. Figure~\ref{fig:EbvsBccvs2c800G}
  shows that for the entrance-channel dominated diatomic zero-energy 
  resonances \cite{Koehler2006a} studied in this paper, $B_\mathrm{res}$ is 
  considerably shifted with respect to the measurable position $B_0$ of the 
  singularity of $a(B)$ referred to in Eq.~(\ref{aofB}).
  
  \subsection{Implementation of the AGS equations}
  \label{subsec:implementation}
  
  \subsubsection{Interchannel coupling}
  
  For convenience, the contributions to Eq.~(\ref{potentials}) describing 
  interchannel coupling can always be represented in terms of an amplitude 
  $\zeta$, and a wave function $|\chi\rangle$, i.e.
  \begin{equation}
    W|\phi_\mathrm{res}\rangle=|\chi\rangle\zeta.
    \label{couplinggen}
  \end{equation}
  Within a limited range of energies close to the dissociation threshold the 
  precise functional form of $\chi(r)=\langle\mathbf{r}|\chi\rangle$, 
  associated with its dependence on the distance $r=|\mathbf{r}|$ between 
  atoms of an interacting pair, is unresolved. As a consequence, several 
  different models of coupling are capable of predicting the same low-energy 
  physics. For this reason, it is sufficient for all applications in this 
  paper to describe the interchannel coupling simply in terms of an amplitude 
  and a range. In accordance with Ref.~\cite{Goral2004a}, we choose the 
  following arbitrary but convenient Gaussian function in momentum space:
  \begin{equation}
    \chi(p)=\langle\mathbf{p}|\chi\rangle=
    \frac{e^{-p^2\sigma^2/(2\hbar^2)}}{(2\pi\hbar)^{3/2}}.
    \label{couplingGauss}
  \end{equation} 
  Here $\sigma$ is the associated range parameter, $\mathbf{p}$ denotes the 
  momentum of the relative motion of an atom pair, and $p=|\mathbf{p}|$ is its
  modulus. The model parameters $\zeta$ and $\sigma$ are determined mainly by 
  the resonance width $\Delta B$ of Eq.~(\ref{aofB}) and the resonance shift 
  $B_0-B_\mathrm{res}$ \cite{Koehler2006a}. Interchannel coupling is also 
  responsible for the avoided crossing of the levels $E_{-1}^\mathrm{d}$ and 
  $E_{-2}^\mathrm{d}$, illustrated in Figs.~\ref{fig:133CsEbvsB} and
  \ref{fig:EbvsBccvs2c800G}. 
  
  \subsubsection{Background scattering}
  
  The background-scattering potential $V_\mathrm{bg}$ describes the 
  interaction of a pair of entrance-channel atoms in the absence of 
  interchannel coupling at magnetic fields asymptotically far away from 
  diatomic zero-energy resonances. In combination with the resonance state, 
  its bare vibrational energy levels determine the number and main properties 
  of comparatively tightly bound dressed, two-channel target states for 
  three-body recombination at negative $a$. Their significance tends to 
  decrease with increasing binding energy. Whereas realistic alkali atom pair 
  potentials typically support dozens of dimer levels, we found it sufficient 
  to account for only the two highest excited vibrational states in our model 
  of $V_\mathrm{bg}$ to determine the correct magnitudes of $K_3(B)$ in 
  Fig.~\ref{fig:133CsK3vsB}. To this end it is crucial, however, to precisely 
  mimic the properties of the dressed dimer energy levels $E_{-1}^\mathrm{d}$ 
  and $E_{-2}^\mathrm{d}$ indicated in Figs.~\ref{fig:133CsEbvsB} and 
  \ref{fig:EbvsBccvs2c800G}, which have been inferred from exact 
  coupled-channels calculations \cite{Chin2004a}. 
  
  Similarly to the interchannel coupling, the precise functional form of the
  microscopic potential $V_\mathrm{bg}(r)$ is unresolved for the energy range
  of interest, leaving a choice of models predicting the same physics. To 
  describe background scattering as well as the two highest excited 
  vibrational levels of $V_\mathrm{bg}(r)$ we employ the following separable 
  representation \cite{Yamaguchi1954a,Mitra1962a} of the background-scattering 
  potential:
  \begin{equation}
    V_\mathrm{bg}=\sum_{j=1}^2 |\chi_j\rangle\xi_j\langle\chi_j|,
    \label{Vbgsep}
  \end{equation}
  Such pseudo-interactions are particularly useful in numerically exact 
  treatments of the three-body Schr\"odinger equation and can be strictly 
  motivated by complete separable expansions of diatomic transition matrices 
  \cite{Sandhas1972a,Lovelace1964a}. In accordance with 
  Ref.~\cite{Goral2004a}, we choose each form factor $|\chi_j\rangle$ of 
  Eq.~(\ref{Vbgsep}) to be a Gaussian function in momentum space:
  \begin{equation}
    \chi_j(p)=\langle\mathbf{p}|\chi_j\rangle=
    \frac{e^{-p^2\sigma_j^2/(2\hbar^2)}}{(2\pi\hbar)^{3/2}}.
    \label{VbgGauss}
  \end{equation}
  The amplitudes $\xi_j$ and range parameters $\sigma_j$ are determined for 
  $j=1,2$ in such a way that Eq.~(\ref{Vbgsep}) recovers $a_\mathrm{bg}$ of 
  Eq.~(\ref{aofB}), the effective range of the microscopic 
  background-scattering potential \cite{Gao1998a,Flambaum1999a}, as well as 
  its two highest excited vibrational energy levels.
  
  Given the resonance state energy $E_\mathrm{res}(B)$ of Eq.~(\ref{EresofB}),
  Eqs.~(\ref{couplinggen}), (\ref{couplingGauss}), (\ref{Vbgsep}) and
  (\ref{VbgGauss}) completely set up our single-resonance model of the pair 
  interaction (\ref{potentials}). We note that the associated diatomic 
  two-channel bound-state energies do not depend on the functional form of 
  $\phi_\mathrm{res}(r)$ \cite{Koehler2006a}, whereas, in general, 
  calculations of $K_3$ loss-rate constants could be sensitive to it. It turns 
  out that the orthogonality assumption of Eq.~(\ref{orthogonalitycl}) allows 
  us to treat three-body recombination without specifying the wave function 
  $\phi_\mathrm{res}(r)$. Similar phenomena have been discussed in the context 
  of pairing in ultracold Fermi gases \cite{BCSpairing}. The predictive power 
  of our model has so far been demonstrated not only in applications to 
  two-body physics \cite{Koehler2006a}, but also to resonance-enhanced 
  three-body decay of $^{87}$Rb Bose-Einstein condensates \cite{Smirne2006a} 
  as well as atom-dimer collisions \cite{Syassen2006a}. In contrast to the 
  broad diatomic $^{133}$Cs and $^{85}$Rb resonances of the present paper, the 
  narrow $^{87}$Rb 1007\,G resonance of Refs.~\cite{Smirne2006a} and 
  \cite{Syassen2006a} is closed-channel dominated \cite{Koehler2006a}.

  \subsubsection{AGS integral equations}

  Given the pair interactions of the three-body Hamiltonian (\ref{H3B}), the 
  general AGS equations~(\ref{AGSpsialpha1}) can be converted into integral 
  equations by multiplying them by ${_\alpha\langle}\mathbf{q},\mathbf{p}|$ 
  from the left and inserting the completeness relation,
  \begin{equation}
    \int d\mathbf{p}\int d\mathbf{p}\,|\mathbf{q},\mathbf{p}\rangle_\gamma
    \,{_\gamma\langle}\mathbf{q},\mathbf{p}|=1,
  \end{equation}
  for each $\gamma\neq\alpha$ into the sum on the right-hand side. Using a 
  partial wave expansion, the numerical solution of Eq.~(\ref{AGSpsialpha1}) 
  is equivalent to matrix inversion on a two-dimensional grid involving both 
  Jacobi momenta $q=|\mathbf{q}|$ and $p=|\mathbf{p}|$. This procedure is 
  commonly known as the momentum space Faddeev approach \cite{Gloeckle1983a}. 
  
  Similarly to the usual separable-potential approach 
  \cite{Sandhas1972a,Lovelace1964a}, it is the particular form of 
  Eq.~(\ref{potentials}) in combination with Eq.~(\ref{orthogonalitycl}) that 
  allow us to reduce this problem to a one-dimensional integral equation in 
  the variable $q$. To this end, we introduce the following complete 
  transition amplitudes, 
  \begin{align}
    \nonumber
    \tilde{X}_{\mathrm{bg},j}(\mathbf{q})&=\frac{1}{3}
    \sum_{\alpha=1}^3 
	{_\alpha\langle}\mathbf{q},\chi_j,\mathrm{bg}|G_\alpha(z)
	|\psi_{\alpha,1}\rangle,\\
	\tilde{X}_{\mathrm{cl}}(\mathbf{q})&=\frac{1}{3}
	\sum_{\alpha=1}^3 
    {_\alpha\langle}\mathbf{q},\phi_\mathrm{res},\mathrm{cl}|G_\alpha(z)
    |\psi_{\alpha,1}\rangle,
    \label{Xprelim}
  \end{align}  
  and inhomogeneous terms,
  \begin{align}
    \nonumber  
    \tilde{X}_{\mathrm{bg},j}^\mathrm{i}(\mathbf{q})&=\frac{1}{3}
    \sum_{\alpha=1}^3 
	{_\alpha\langle}\mathbf{q},\chi_j,\mathrm{bg}|G_\alpha(z)
	|\psi_{\alpha,1}^\mathrm{i}\rangle,\\
	\tilde{X}_{\mathrm{cl}}^\mathrm{i}(\mathbf{q})&=\frac{1}{3}
	\sum_{\alpha=1}^3 
	    {_\alpha\langle}\mathbf{q},\phi_\mathrm{res},\mathrm{cl}|
	    G_\alpha(z)
	    |\psi_{\alpha,1}^\mathrm{i}\rangle.
     \label{Xinhprelim}
  \end{align}
  Here the summation over the index $\alpha=1,2,3$ refers to the equivalence 
  of particle arrangements due to the identical nature of the three Bose 
  atoms. Given these amplitudes, multiplying the AGS 
  equations~(\ref{AGSpsialpha1}) by 
  ${_\alpha\langle}\mathbf{q},\chi_j,\mathrm{bg}|G_\alpha(z)$ and 
  ${_\alpha\langle}\mathbf{q},\phi_\mathrm{res},\mathrm{cl}|G_\alpha(z)$ from 
  the left and summing over the index $\alpha$ yields the following integral 
  equations:
  \begin{align}
    \nonumber
    \tilde{X}_{\mathrm{bg},j}(\mathbf{q})=&
    \tilde{X}_{\mathrm{bg},j}^\mathrm{i}(\mathbf{q})+
    \sum_{k=1}^2 \int d\mathbf{q}'\, 
    \tilde{\mathcal{K}}_{jk}^{\mathrm{bg},\mathrm{bg}}
    (\mathbf{q},\mathbf{q}')
    \tilde{X}_{\mathrm{bg},k}(\mathbf{q}')\\
    \nonumber
    &+\int d\mathbf{q}'\, 
    \tilde{\mathcal{K}}_{j}^{\mathrm{bg},\mathrm{cl}}
    (\mathbf{q},\mathbf{q}')
    \tilde{X}_{\mathrm{cl}}(\mathbf{q}'),\\
    \nonumber
    \tilde{X}_{\mathrm{cl}}(\mathbf{q})=&
    \tilde{X}_{\mathrm{cl}}^\mathrm{i}(\mathbf{q})+
    \sum_{k=1}^2 \int d\mathbf{q}'\, 
    \tilde{\mathcal{K}}_{k}^{\mathrm{cl},\mathrm{bg}}
    (\mathbf{q},\mathbf{q}')
    \tilde{X}_{\mathrm{bg},k}(\mathbf{q}')\\
    &+\int d\mathbf{q}'\,
    \tilde{\mathcal{K}}_{\mathrm{cl},\mathrm{cl}}
    (\mathbf{q},\mathbf{q}')
    \tilde{X}_{\mathrm{cl}}(\mathbf{q}').
    \label{AGSintprelim}
  \end{align}
  Here the index $j$ assumes the values 1 and 2 and the integral kernels are 
  given by the following expressions: 
  \begin{align}
    \nonumber
    \tilde{\mathcal{K}}_{jk}^{\mathrm{bg},\mathrm{bg}}
    (\mathbf{q},\mathbf{q}')=&
    2\,{_2\langle}\mathbf{q},\chi_j,\mathrm{bg}|G_2(z)
    |\mathbf{q}',\chi_k,\mathrm{bg}\rangle_1\,\xi_k,\\
    \nonumber
    \tilde{\mathcal{K}}_{j}^{\mathrm{bg},\mathrm{cl}}
    (\mathbf{q},\mathbf{q}')=&
    2\,{_2\langle}\mathbf{q},\chi_j,\mathrm{bg}|G_2(z)
    |\mathbf{q}',\chi,\mathrm{bg}\rangle_1\,\zeta,\\
    \nonumber
    \tilde{\mathcal{K}}_{k}^{\mathrm{cl},\mathrm{bg}}
    (\mathbf{q},\mathbf{q}')=&
    2\,{_2\langle}\mathbf{q},\phi_\mathrm{res},\mathrm{cl}|G_2(z)
    |\mathbf{q}',\chi_k,\mathrm{bg}\rangle_1\,\xi_k,\\
    \tilde{\mathcal{K}}_{\mathrm{cl},\mathrm{cl}}
    (\mathbf{q},\mathbf{q}')=&
    2\,{_2\langle}\mathbf{q},\phi_\mathrm{res},\mathrm{cl}|G_2(z)
    |\mathbf{q}',\chi,\mathrm{bg}\rangle_1\,\zeta.
    \label{kernelsprelim}
  \end{align}
  The indices $\alpha=2$ and $\gamma=1$ of the final and initial particle 
  arrangements in these integral kernels are arbitrary as long as they differ, 
  and the prefactors of 2 stem from the summation over indices 
  $\gamma\neq\alpha$ in the general AGS equations~(\ref{AGSpsialpha1}). As the 
  kernels (\ref{kernelsprelim}) do not couple different partial waves
  associated with $\mathbf{q}$, Eqs.~(\ref{AGSintprelim}) can be treated, in 
  general, as independent sets of integral equations in the variable 
  $q=|\mathbf{q}|$ for each partial wave individually. Since we have 
  restricted ourselves to zero-energy initial conditions ($z=i0$) the 
  inhomogeneous terms are isotropic, 
  i.e.~$\tilde{X}_{\mathrm{bg},j}^\mathrm{i}(\mathbf{q})=
  \tilde{X}_{\mathrm{bg},j}^\mathrm{i}(q)$ and
  $\tilde{X}_{\mathrm{cl}}^\mathrm{i}(\mathbf{q})=
  \tilde{X}_{\mathrm{cl}}^\mathrm{i}(q)$. Consequently, for all applications 
  in this paper Eqs.~(\ref{AGSintprelim}) involve the $s$-wave only. 
  
  \subsubsection{Numerical treatment of the AGS integral equations}
  
  Due to the arrangement-channel Green's function, $G_2(z)$, the kernels of 
  Eqs.~(\ref{kernelsprelim}) have poles at the physical energies associated 
  with dressed, coupled-channels bound states, as indicated in the upper panel 
  of Fig.~\ref{fig:133CsEbvsB}. In order to explicitly locate the positions of 
  these singularities, we introduce the effective diatomic energy variable of 
  the relative motion, $E=z-3q^2/(4m)$, as well as the entrance-channel 
  Green's function,  
  \begin{equation}
    \hat{G}_\mathrm{bg}(E)=
    \left(
      E+\hbar^2\boldsymbol{\nabla}^2/m-V_\mathrm{bg}
      \right)^{-1}.
    \label{Gbg}
  \end{equation}
  According to Ref.~\cite{Goral2004a}, the resonance denominator of the 
  complete diatomic Green's function $G_2(z)$ is given by  
  \begin{equation}
    \mathcal{E}(B,E)=E-E_\mathrm{res}(B)-|\zeta|^2
    \langle\chi|\hat{G}_\mathrm{bg}(E)|\chi\rangle.
    \label{resdenom}
  \end{equation}
  At each magnetic-field strength $B$, its zeros determine the locations of 
  all diatomic bound-state energies $E_\mathrm{d}=E_\mathrm{d}(B)$ through 
  $\mathcal{E}(B,E_\mathrm{d})=0$. The dressed, coupled-channels dimer states 
  are characterized by the wave-function normalization coefficient, 
  \begin{equation}
    Z(B,E)=\frac{1}{\frac{\partial}{\partial E}\mathcal{E}(B,E)},
    \label{ZofBE}
  \end{equation}
  which determines their closed-channel admixtures in the lower panel of 
  Fig.~\ref{fig:133CsEbvsB} via the relation $Z(B)=Z(B,E_\mathrm{d})$. This 
  coefficient in combination with the arrangement-channel Green's function and 
  its resonance denominator determine the final state of 
  Eq.~(\ref{AGSamplitudepsialpha1}) to be
  \begin{equation}
    |\mathbf{q},\phi_\mathrm{d}\rangle_\alpha=
    G_\alpha(z)
    |\mathbf{q},\phi_\mathrm{res},\mathrm{cl}\rangle_\alpha\,
    \mathcal{E}(B,E)\sqrt{Z(B,E)}
  \end{equation}
  in the limit $E\to E_\mathrm{d}(B)$. 
  
  Besides the poles in Eqs.~(\ref{Xinhprelim}) and (\ref{kernelsprelim}) at 
  the diatomic bound-state energies, the resonance denominator of 
  Eq.~(\ref{resdenom}) as well as the coefficient of Eq.~(\ref{ZofBE}) involve 
  singularities due to the bare Green's function of Eq.~(\ref{Gbg}). Although 
  these unphysical singularities associated with bare vibrational levels 
  cancel exactly in the AGS integral equations~(\ref{AGSintprelim}), for any 
  numerical implementation of Eqs.~(\ref{AGSintprelim}) it is imperative to 
  explicitly treat these cancellations. To this end, we introduce the 
  modified, dimensionless transition amplitudes, 
  \begin{align}
    \nonumber
    X_{\mathrm{bg},j}(q)&=q\tilde{X}_{\mathrm{bg},j}(q)
    \frac{\mathcal{E}(B,E)}{\sqrt{\langle\chi_j|\chi_j\rangle}} 
    \sqrt{Z(B,E)},\\
    X_{\mathrm{cl}}(q)&=q\tilde{X}_{\mathrm{cl}}(q)
    \mathcal{E}(B,E)\sqrt{Z(B,E)},
    \label{Xfinal}
  \end{align}  
  in addition to the following inhomogeneous terms:   
  \begin{align}
    \nonumber  
    X_{\mathrm{bg},j}^\mathrm{i}(q)&=q\tilde{X}_{\mathrm{bg},j}^\mathrm{i}(q)
    \frac{\mathcal{E}(B,E)}{\sqrt{\langle\chi_j|\chi_j\rangle}}
    \sqrt{Z(B,E)},\\
    X_{\mathrm{cl}}^\mathrm{i}(q)&=q\tilde{X}_{\mathrm{cl}}^\mathrm{i}(q)
    \mathcal{E}(B,E)\sqrt{Z(B,E)}.  
  \end{align}
  Here the index $j$ assumes the values 1 and 2, similarly to 
  Eqs.~(\ref{Xprelim}) and (\ref{Xinhprelim}). The associated AGS integral 
  equations can be readily obtained from Eqs.~(\ref{AGSintprelim}), which 
  yields:
  \begin{align}
    \nonumber
    X_{\mathrm{bg},j}(q)=&
    X_{\mathrm{bg},j}^\mathrm{i}(q)+
    \sum_{k=1}^2 \int_0^\infty dq'\, 
    \mathcal{K}_{jk}^{\mathrm{bg},\mathrm{bg}}(q,q')
    X_{\mathrm{bg},k}(q')\\
    \nonumber
    &+\int_0^\infty dq'\, 
    \mathcal{K}_{j}^{\mathrm{bg},\mathrm{cl}}(q,q')
    X_{\mathrm{cl}}(q'),\\
    \nonumber
    X_{\mathrm{cl}}(q)=&
    X_{\mathrm{cl}}^\mathrm{i}(q)+
    \sum_{k=1}^2 \int_0^\infty dq'\, 
    \mathcal{K}_{k}^{\mathrm{cl},\mathrm{bg}}(q,q')
    X_{\mathrm{bg},k}(q')\\
    &+\int_0^\infty dq'\, 
    \mathcal{K}_{\mathrm{cl},\mathrm{cl}}(q,q')
    X_{\mathrm{cl}}(q').
    \label{AGSint}
  \end{align}
  Here the modified kernels are given in terms of Eqs.~(\ref{kernelsprelim}) 
  by the following expressions:
  \begin{align}
    \nonumber
    \mathcal{K}_{jk}^{\mathrm{bg},\mathrm{bg}}(q,q')=&
    qq'
    \sqrt{\frac{Z(B,E)}{Z(B,E')}}
    \frac{\mathcal{E}(B,E)}{\mathcal{E}(B,E')}
    \xi_k\sqrt{\frac{\langle\chi_k|\chi_k\rangle}
      {\langle\chi_j|\chi_j\rangle}}\\
    \nonumber
    &\times
    \int d\Omega'\,
    \tilde{\mathcal{K}}_{jk}^{\mathrm{bg},\mathrm{bg}}
    (\mathbf{q},\mathbf{q}'),
    \\
    \nonumber
    \mathcal{K}_{j}^{\mathrm{bg},\mathrm{cl}}(q,q')=&
    qq'
    \sqrt{\frac{Z(B,E)}{Z(B,E')}}
    \frac{\mathcal{E}(B,E)}{\mathcal{E}(B,E')}
    \frac{\zeta}{\sqrt{\langle\chi_j|\chi_j\rangle}}\\
    \nonumber
    &\times
    \int d\Omega'\,
    \tilde{\mathcal{K}}_{j}^{\mathrm{bg},\mathrm{cl}}
    (\mathbf{q},\mathbf{q}'),
    \\
    \nonumber
    \mathcal{K}_{k}^{\mathrm{cl},\mathrm{bg}}(q,q')=&
    qq'
    \sqrt{\frac{Z(B,E)}{Z(B,E')}}
    \frac{\mathcal{E}(B,E)}{\mathcal{E}(B,E')}
    \xi_k\sqrt{\langle\chi_k|\chi_k\rangle}\\
    \nonumber
    &\times
    \int d\Omega'\,
    \tilde{\mathcal{K}}_{k}^{\mathrm{cl},\mathrm{bg}}
    (\mathbf{q},\mathbf{q}'),
    \\
    \nonumber
    \mathcal{K}_{\mathrm{cl},\mathrm{cl}}(q,q')=&
    qq'
    \sqrt{\frac{Z(B,E)}{Z(B,E')}}
    \frac{\mathcal{E}(B,E)}{\mathcal{E}(B,E')}
    \zeta\\
    &\times
    \int d\Omega'\,
    \tilde{\mathcal{K}}_{\mathrm{cl},\mathrm{cl}}
    (\mathbf{q},\mathbf{q}').
  \end{align}
  Here $d\Omega'$ denotes the angular component of $d\mathbf{q}'$ and stems
  from the momentum integrals of Eqs.~(\ref{AGSintprelim}).
  
  The transformed one-dimensional AGS integral equations~(\ref{AGSint}) are 
  numerically stable and can be readily solved for the complete transition
  amplitudes of Eqs.~(\ref{Xfinal}). Using Eqs.~(\ref{K3thermal}) and 
  (\ref{AGSamplitudepsialpha1}), their solution determines the three-body 
  recombination loss-rate constant via the following relation:  
  \begin{equation}
    K_3=
    48\pi^2(2\pi\hbar)^6m\frac{q_\mathrm{f}}{\hbar}
    \left|
      \frac{X_\mathrm{cl}(q_\mathrm{f})}{q_\mathrm{f}}
      \right|^2.
  \end{equation}
  Here the final Jacobi momentum $q_\mathrm{f}$ is given by Eq.~(\ref{qf}). 
  Our implementation of the AGS method was used to determine the 
  magnetic-field dependence of the $K_3$ loss-rate constants of 
  Ref.~\cite{Smirne2006a} as well as in the upper panels of 
  Figs.~\ref{fig:133CsK3vsB} and \ref{fig:133CsK3vsB800G}. A similar treatment 
  using the same kernels but different inhomogeneous terms has been used to 
  predict the scattering amplitudes for atom-dimer collisions in ultracold 
  Bose gases of $^{87}$Rb \cite{Smirne2006a}. 
  
  \subsection{Parameters for $^{133}$Cs Fesh\-bach resonances}
  
  \subsubsection{Diatomic entrance and closed channels}
  
  The nuclear spin quantum number $I=7/2$ of $^{133}$Cs \cite{Arimondo1977a} 
  gives rise to two hyperfine manifolds labeled by the total atomic angular 
  momentum quantum numbers, $F=3$ and $F=4$. Due to the magnetic field, each 
  hyperfine level is split into Zeeman sublevels. Interatomic collisions in an 
  ultracold gas are sensitive to the Zeeman state in which the atoms are 
  prepared. Throughout this paper, this Zeeman state is characterized by the 
  $(F=3,m_F=+3)$ angular momentum quantum numbers of the atomic state which it 
  adiabatically correlates with in the limit of zero field. Due to the 
  spatially homogeneous nature of the magnetic field, all pair potentials are 
  invariant with respect to rotations about the field axis. Both the high- and 
  low-field $^{133}$Cs Fesh\-bach resonances studied in this paper interact 
  with the diatomic entrance-spin-channel mainly via spin exchange 
  \cite{Stoof1988a,Tiesinga1993a}. These interactions do not couple different 
  partial waves associated with the relative motion of an atom pair. 
  Consequently, the entrance-spin-channel, $(F_1,m_1;F_2,m_2)=(3,3;3,3)$, 
  associated with a pair of $^{133}$Cs atoms 1 and 2 is coupled only to those 
  Zeeman states which conserve the sum of single-particle angular momentum 
  projection quantum numbers, $m_1+m_2=6$. All possible strongly coupled 
  closed channels are therefore characterized by the pairs of atomic quantum 
  numbers $(3,2;4,4)$, $(3,3;4,3)$, $(4,3;4,3)$, and $(4,4;4,2)$. Their 
  associated magnetic moments can be estimated by the Breit-Rabi formula 
  \cite{Breit1931a} to determine the bare resonance-state energies 
  $E_\mathrm{res}(B)$ of Eq.~(\ref{EresofB}) \cite{Koehler2006a}.
  
  \subsubsection{Low-field zero-energy resonance}
  
  The low-field zero-energy resonance at the negative magnetic-field strength 
  \cite{Vogels1998a} of $B_0=-11.2$\,G is caused by a Fesh\-bach resonance 
  level, $E_\mathrm{res}(B)$, comprised of Zeeman states mainly from the 
  excited $F=4$ hyperfine level. Independently of the specific spin 
  composition of the associated strongly coupled closed-spin-channel state the 
  Breit-Rabi formula yields $\mu_\mathrm{res}=h\times 4.2\,$MHz/G for the 
  magnetic moment difference of Eq.~(\ref{EresofB}). Coupled-channels 
  scattering calculations of the resonance-enhanced scattering length $a(B)$ 
  including basis states of both $s$-wave and $d$-wave symmetries determine 
  the parameters $a_\mathrm{bg}$ and $\Delta B$ given in 
  Subsection~\ref{subsec:lowfield}. Besides $a_\mathrm{bg}$, the energies of 
  the two highest excited bare vibrational entrance-channel states of 
  $-h\times 0.01\,$MHz and $-h\times 111\,$MHz, as well as the effective range 
  $r_\mathrm{eff}=252\,a_0$ determine the separable background-scattering 
  potential of Eq.~(\ref{Vbgsep}). Given the background-scattering length, 
  both of the bare energies \cite{Gao1998a} as well as $r_\mathrm{eff}$ 
  \cite{Gao1998a,Flambaum1999a} are determined mainly by the van der Waals 
  dispersion coefficient $C_6=6890\,$a.u.~of cesium \cite{Leo2000a} (the 
  atomic unit of $C_6$ is $1\,$a.u.$=9.5734\times10^{-26}\,$J\,nm$^6$). The 
  resonance width $\Delta B$ and the shift $B_0-B_\mathrm{res}=-29\,$G 
  \cite{Koehler2006a} determine our model of the interchannel coupling. All 
  these physical parameters give the model parameters of 
  Eqs.~(\ref{couplinggen}), (\ref{couplingGauss}), (\ref{Vbgsep}), and 
  (\ref{VbgGauss}) to be: 
  $\xi_1=-5.4187\times 4\pi^{3/2}\hbar^2\sigma_1/m$, 
  $\xi_2=-12.9603\times 4\pi^{3/2}\hbar^2\sigma_2/m$, 
  $\zeta=0.59623807\times 2\pi^{3/4}\hbar^2/(m\sigma^{1/2})$, 
  $\sigma_1=82.3601\,a_0$, $\sigma_2=36.4963\,a_0$, and $\sigma=14.4694\,a_0$.
  
  \subsubsection{High-field zero-energy resonance}

  The high-field zero-energy resonance at $B_0=800\,$G exists due to strong 
  coupling between the diatomic $(3,3;3,3)$ entrance-spin-channel and a closed 
  spin channel comprised of the $(3,2;4,4)$ and $(3,3;4,3)$ Zeeman-state 
  configurations. According to coupled-channels calculations of the $s$-wave 
  bound states, their relative admixtures to the closed-channel superposition 
  state are about 57\% and 43\%, respectively. Based on the Breit-Rabi 
  formula, these admixtures imply an estimated magnetic-moment difference 
  associated with the Fesh\-bach-resonance state of 
  $\mu_\mathrm{res}=h\times2.6\,$MHz/G. We note that due to the 
  single-particle spin overlap between the entrance and closed channels, 
  Eq.~(\ref{orthogonalitycl}) does not strictly apply to the $800\,$G 
  high-field zero-energy resonance. We believe, however, that the associated
  sensitivity to short-distance physics, such as the functional form of the
  resonance state, $\phi_\mathrm{res}(r)$, is sufficiently small for our 
  calculations to predict the correct orders of magnitude of $K_3$. Our 
  estimates of three-body zero-energy resonance positions are unaffected by
  spin-overlap, as they only involve energy scales much closer to the 
  dissociation threshold than three-body recombination into tightly bound 
  diatomic target states at negative $a(B)$.
  
  The physical parameters $a_\mathrm{bg}$ and $\Delta B$ determining the 
  resonance-enhanced scattering length of Eq.~(\ref{aofB}) near 800\,G are 
  given in Subsection~\ref{subsec:cesium800G}. Based on $a_\mathrm{bg}$ and 
  $C_6$ we calculate the energies of the two highest excited vibrational 
  levels of the bare background-scattering potential to be 
  $-h\times 0.0086\,$MHz and $-h\times 110\,$MHz, while the effective range
  and resonance shift amount to $r_\mathrm{eff}=255\,a_0$ and 
  $B_0-B_\mathrm{res}=-87\,$G, respectively. Accordingly, the model parameters 
  of Eqs.~(\ref{couplinggen}), (\ref{couplingGauss}), (\ref{Vbgsep}), and 
  (\ref{VbgGauss}) for the high-field zero-energy resonance are given by: 
  $\xi_1=-6.5718\times 4\pi^{3/2}\hbar^2\sigma_1/m$, 
  $\xi_2=-13.7782\times 4\pi^{3/2}\hbar^2\sigma_2/m$, 
  $\zeta=0.92953214\times 2\pi^{3/4}\hbar^2/(m\sigma^{1/2})$, 
  $\sigma_1=80.6241\,a_0$, $\sigma_2=38.512\,a_0$, and $\sigma=18.7179\,a_0$.

\end{document}